# RECOMBINATION LINE VS. FORBIDDEN LINE ABUNDANCES IN PLANETARY NEBULAE

MARK ROBERTSON-TESSI
Steward Observatory, University of Arizona, 933 North Cherry Avenue, Tucson, AZ 85721; mro@email.arizona.edu

AND

DONALD R. GARNETT
Steward Observatory, University of Arizona, 933 North Cherry Avenue, Tucson, AZ 85721; dgarnett@as.arizona.edu

## ABSTRACT

Recombination lines (RLs) of C II, N II, and O II in planetary nebulae (PNs) have been found to give abundances that are much larger in some cases than abundances from collisionally-excited forbidden lines (CELs). The origins of this abundance discrepancy are highly debated. We present new spectroscopic observations of O II and C II recombination lines for six planetary nebulae. With these data we compare the abundances derived from the optical recombination lines with those determined from collisionally-excited lines. Combining our new data with published results on RLs in other PNs, we examine the discrepancy in abundances derived from RLs and CELs. We find that there is a wide range in the measured abundance discrepancy $\Delta(O^{+2})$ = log $O^{+2}$(RL) - log $O^{+2}$(CEL), ranging from approximately 0.1 dex (within the 1$\sigma$ measurement errors) up to 1.4 dex. This tends to rule out errors in the recombination coefficients as a source of the discrepancy. Most RLs yield similar abundances, with the notable exception of O II multiplet V15, known to arise primarily from dielectronic recombination, which gives abundances averaging 0.6 dex higher than other O II RLs. We compare $\Delta(O^{+2})$ against a variety of physical properties of the PNs to look for clues as to the mechanism responsible for the abundance discrepancy. The strongest correlations are found with the nebula diameter and the Balmer surface brightness; high surface brightness, compact PNs show small values of $\Delta(O^{+2})$, while large low surface brightness PNs show the largest discrepancies. An inverse correlation of $\Delta(O^{+2})$ with nebular density is also seen. A marginal correlation of $\Delta(O^{+2})$ is found with expansion velocity. No correlations are seen with electron temperature, $He^{+2}/He^{+}$, central star effective temperature and luminosity, stellar mass loss rate, or nebular morphology. Similar results are found for carbon in comparing C II RL abundances with ultraviolet measurements of C III].

## 1. INTRODUCTION

The abundances of heavy elements in planetary nebulae (PNs) are important for understanding their evolution. PNs often show enrichments (or depletions) of He, N, and C that point to the effects of nuclear reactions deep within the asymptotic giant branch (AGB) progenitor star; the relative enrichments are a clue to the mass of the progenitor, as the amounts of C and N depend on the mass of the star. On a collective scale, abundance measurements for elements that are not affected by nuclear processing in PNs (such as O, Ne, S , or Ar) can be used to trace the variation of metallicity across the disk of the Galaxy, and even provide some information on the time evolution of abundances and abundance gradients (Maciel et al. 2003).

Ion and element abundances in PNs are most often determined from measurements of the intensities of forbidden lines excited by electron impact excitation. For the more abundant elements these collisionally-excited lines (CELs) are bright and can be accurately measured; if one can estimate electron temperatures and electron densities from certain diagnostic line ratios, then it is possible to derive element abundances (relative to hydrogen) fairly accurately. However, for CELs in the visible and ultraviolet parts of the spectrum, the abundance determined can be very sensitive to errors in the determination of the electron temperature of the nebula. CELs in the infrared have much weaker temperature dependence, but important emission lines such as [O III] 52, 88 microns and the 57 micron line of [N III] are sensitive to density variations. Thus, it has been suggested that optical recombination lines (RLs) from heavy elements may provide more accurate abundances than CELs:  RLs from heavy elements have similar temperature and density dependences as the familiar hydrogen Balmer lines, so abundances derived from RLs are nearly independent of electron temperature. Nevertheless, RLs are typically more than 100 times fainter than the $H\beta$ line, and thus are much more difficult to detect and measure.

Early studies with older instrumentation suggested that recombination lines for C II and O II gave larger abundances than corresponding collisionally-excited C III] and [O III] lines (Kaler 1986; Barker 1991; Peimbert, Storey, & Torres-Peimbert 1993), but close examination showed that large observational uncertainties precluded definite conclusions (Rola & Stasinska 1994). More recent studies have measured RLs in planetary nebulae with greatly improved sensitivity (Liu et al. 1995, 2000, 2001; Garnett & Dinerstein 2001; Tsamis et al. 2003, 2004). These studies have found that ion abundances derived from RLs and CELs are clearly discrepant: RLs usually give higher abundances. The discrepancy can be very large, as much as a factor of 20. The reasons for the discrepancy are highly debated. Peimbert et al. (1993) suggested that local temperature fluctuations (Peimbert 1967) could be the source. Liu et al. (2000) has argued that temperature fluctuations cannot account simultaneously for the large discrepancy between abundances from RLs and optical CELs and the small difference between abundances derived from optical and IR CELs. Liu et al. (2000) found that they could account for the RL-CEL discrepancy in NGC 6153 if the RLs arise mainly in dense, hydrogen-poor clumps with a small filling factor, while the CELs are produced in the lower-density photoionized material. Meanwhile, Liu et al. (2000) and Garnett & Dinerstein (2001) found that O II RLs in NGC 6153 and NGC 6720 were much more centrally concentrated than the [O III] emission. Barker (1991) found a roughly similar relationship between C II and C III] in NGC 2392. Garnett & Dinerstein (2001) suggested that high-temperature dielectronic recombination, in a high temperature gas bubble caused by the interaction between the fast stellar wind and the nebular material, could lead to enhancement of the recombination lines. Mathis (1996) discusses a variety of possible explanations for the RL-CEL abundance discrepancy. None of the proposed mechanisms provides a fully satisfactory explanation at the present time.

Despite the recent increase in activity on RLs, measurements have been made for only a few objects. A larger sample of measurements can provide a valuable statistical sample, in which it would be possible to compare the observed abundance discrepancies with nebular properties to look for clues pointing toward the most likely physical cause of the RL-CEL abundance differences. Here we present new spectral measurements of optical recombination lines in six PNs, and compare the results with other studies. In combination with published data for other planetary nebulae, we compare the observed discrepancies against various physical properties, which can offer clues as to the physical mechanism behind the discrepancies.

## 2. OBSERVATIONS AND MEASUREMENTS

### 2.1. *Observations and Data Reduction*

The spectra were all obtained with the 2.3m Bok reflector plus B&C spectrograph of Steward Observatory on the nights of 7 and 8 June 1999. Both nights were clear. We used the 832 line/mm grating in 2nd order with a 1200x800 pixel Loral CCD to observe the spectral region 4150-5000 Å at 0.71 Å per pixel. With a 2.5 arcsec slit, the spectral resolution ranged from 1.9 to 2.2 Å at FWHM, sufficient to resolve many of the brighter recombination lines; the spectral focus was optimized to give the best resolution around 4650 Å.

For this project, we chose planetary nebulae with high surface brightnesses. This meant that most of the nebulae were relatively compact; the nebular diameters generally ranged from 3-20 arcsec, except for NGC 6720, the observations for which have already been published (Garnett & Dinerstein 2001). Our six target PNs and some of their physical properties are listed in Table 1, as well as 16 other PNs from the literature and their properties. Short (60-300 s) and long (3-7 ks) exposures were taken for each PN. The long exposures were designed to provide high signal/noise for the faint O II recombination lines, while the short exposures were made to measure the fluxes for the brightest lines ([O III] λ4959, and in some cases *Hβ*), which were often saturated in the long exposures. The 240″ long slit was oriented along the E-W direction for all of the observations. In most cases, the slit was positioned at the location of the central star or the center of the nebula, although for some of the more extended nebulae we made a second observation at a position offset north or south of the central star. To determine the flux calibration, we observed the spectrophotometric standard stars Feige 66, HZ 44, PG1708+602, and BD+28 4211 (Massey et al. 1988) during the course of each night.

Reduction of the spectra followed common procedures. We subtracted the DC offset and bias from each frame using the CCD overscan and a sequence of zero-length exposures, combined to produce a single bias frame. Pixel-to-pixel sensitivity variations and vignetting along the slit were removed by dividing each exposure by a flat-field image constructed from a combination of high signal/noise exposures of a quartz lamp and the twilight sky. We derived the wavelength scale from observations of a He-Ar lamp; the rms dispersion from a cubic fit to the positions was of order 0.04 Å. We converted the counts from the CCD image to flux units using the sensitivity functions derived for each night of observation from the standard star spectra, correcting for atmospheric extinction using the mean KPNO extinction coefficients. A 2nd-order cubic spline fit to the sensitivity data gave uncorrelated residuals with an rms uncertainty of about 0.02 mag. The calibrated two-dimensional PN images were then collapsed to one-dimensional spectra by summing the data within a window that included only PN emission; the sky background was determined from portions of the image away from the PN and then subtracted.

## 2.2. *Emission Line Measurements*

We measured fluxes for the emission lines using SPLOT in IRAF[1]. Measurements are shown in Table 2. Column 1 shows the observed wavelength, columns 2, 3, and 4 show the laboratory wavelength, identification, and multiplet of the line, and columns 5 and 6 are the observed and reddening corrected fluxes (c.f. section 2.3), both normalized to $I(H\beta)$=100. Column 7 shows the error associated with the measurement; an A indicates an error of less than 5%, B is an error of 5-10%, C is 10-20%, D is 20-30%, E is 30-50%, and F denotes a 2 sigma limit. The error accounts for flat-field uncertainty (about 1%), uncertainty in reddening parameters (about 2%, c.f. section 2.3), and statistical errors of measurement. The statistical error is a combination of the continuum and emission line uncertainty. The statistical error in the line flux $\sigma_l$, was estimated using

$$\sigma_l = \sigma_c \sqrt{N + e/\Delta} \quad (1)$$

(Pérez-Montero & Díaz 2003), where $\sigma_c$ is the RMS uncertainty in the local continuum, $N$ is the width of our measurement in pixels (17 pixels), $e$ is the equivalent width of the line measurement, and $\Delta$ is the dispersion of our spectra (0.71 Å/pixel).

Short exposures were used for the measurement of the brightest lines, for example [O III] λ4959, which is saturated in the long exposures. Individual resolved lines were measured using Gaussian profiles, assuming a linear local continuum under the line profile. Gaussian fitting differed by only a few percent from flux integration and Voigt profile fitting. Partially blended lines were deconvolved using model profiles with fixed FWHM and the wavelength and flux as free parameters. Lines that were within about 2Å of each other were not deconvolved reliably using Gaussian profiling due to the FWHM of our spectra. These emission features were measured using flux integration; for some combined features, theoretical ratios from other resolved lines of the same element could be used to extract the fluxes. For example, [Ar IV] λ4711 is blended with He I λ4713. We subtracted the λ4713 contribution to the [Ar IV] feature using the flux from He I λ4471, assuming the case B λ4713/λ4471 ratio. The ratio depends on both $T_e$ and $n_e$ because of collisional excitation. We used the calculations of Benjamin, Skillman & Smits (1999) to estimate the λ4713/λ4471 ratio, assuming $T_e$ derived from [O III] and $n_e$ derived from [Ar IV]. The λ4713/λ4471 ratio and density were iterated until convergence was reached. Other features for which such theoretical ratios are not known reliably were left as combined fluxes in Table 2, indicated by an asterisk for each wavelength contributing to the feature.

---

[1] IRAF is distributed by the National Optical Astronomy Observatories, which are operated by the Association of Universities for Research in Astronomy, Inc., under cooperative agreement with the National Science Foundation.

### 2.3. *Interstellar Reddening*

The interstellar reddening for each nebula was estimated from the *Hγ/Hβ* ratio. The relationship between observed and intrinsic intensities is

$$\frac{I_{\mathrm{obs}}(H\gamma)}{I_{\mathrm{obs}}(H\beta)} = \frac{I_{\mathrm{int}}(H\gamma)}{I_{\mathrm{int}}(H\beta)} 10^{-c(H\beta)[f(H\gamma)-f(H\beta)]} \, , \quad (2)$$

where $[f(H\gamma)-f(H\beta)] = 0.15$ from the reddening curve in Osterbrock (1989), and $c(H\beta)$ is the logarithmic extinction of *Hβ*. The intrinsic ratio is mildly dependent on temperature: for $T_e$ between 7500 and 15000 K, the ratio varies from 0.465 to 0.473 (Osterbrock 1989). The value of $c(H\beta)$ was used to correct all other observed fluxes according to

$$I_{\mathrm{corr}}(\lambda) = I_{\mathrm{obs}}(\lambda) 10^{c(H\beta)[f(\lambda)-f(H\beta)]} \; . \quad (3)$$

Table 1 shows the values of $c(H\beta)$ for each nebula.

The uncertainty in $c(H\beta)$ has only a modest effect on the reddening corrected line ratios over this short spectral range; the maximum uncertainty in I(λ4150)/I(λ5000) (end to end) is about 2% for our sample of nebulae.

## 3. Data Analysis

### 3.1. *Electron Density and Temperature*

The electron densities, $n_e$, for the nebulae were determined using the [Ar IV] λ4711/λ4740 ratio. [Ar IV] is a $p^3$ ion, and thus exhibits two closely spaced energy levels, $^2D_{3/2}$ and $^2D_{5/2}$, from which transitions to $^4S_{3/2}$ occur. The two $^2D$ levels are close in energy, so that there is little temperature dependence; the ratio of these two transitions is density sensitive because they have different collisional excitation cross-sections. The electron densities for the nebulae in our sample were calculated using the *nebular.temden* software package (Shaw & Dufour 1995) in IRAF. The calculated values for $n_e$ are shown in Table 3, along with results obtained by other studies. For IC 4593, the strength of the [Ar IV] lines was too weak to measure the density accurately.

The electron temperatures, $T_e$, were determined using the [O III] λ4959/λ4363 ratio, which is useful as a temperature diagnostic because the two lines have very different excitation energies. We used the five-level atomic code *nebular.temden* to calculate temperatures for each nebula. The values for $T_e$ are shown in Table 3, along with results obtained by other studies for comparison. The values for $c(H\beta)$, $n_e$, $T_e$, and intrinsic line ratios are all interdependent, so calculations were reiterated until a convergence was reached. In practice, the results converged rapidly.

## 3.2. *Ionic Abundances*

Because of their weak temperature and density dependence, optical recombination lines can be used to provide abundance determinations which are not significantly affected by small scale fluctuations of electron temperature. However, their line emission strengths are usually fairly faint, and thus are prone to higher statistical and systematic measurement error. Collisionally excited lines are usually bright and easily measured, but their abundance determination is highly temperature dependent.

For RLs, the ratio of the intensity of an emission line to the intensity of $H\beta$ is used to calculate the abundance of the ion relative to $H^+$ by

$$\frac{N(X^{+n})}{N(H^{+})} = \frac{\lambda(X)}{\lambda(H\beta)} \frac{I(\lambda)}{I(H\beta)} \frac{\alpha_{\text{eff}}(H\beta)}{\alpha_{\text{eff}}(\lambda)} , \tag{4}$$

where $\alpha_{\text{eff}}(\lambda)$ is the effective recombination coefficient for the transition producing the line $\lambda$. The recombination coefficients are temperature sensitive, proportional to $T^{-m}$, where $m$ is approximately 1 for most ions (Osterbrock 1989). When taken in ratio, the temperature dependence is small, and typically there is less than 5% variation from 10,000K to 15,000K in the ratio of coefficients (Garnett & Dinerstein 2001). The value used for $\alpha_{\text{eff}}(H\beta)$ is from Osterbrock (1989). For He I, $\alpha_{\text{eff}}(\lambda)$ values are from Benjamin et al. (1999), and for He II, the values are from Hummer and Storey (1998). The values for O II are from Storey (1994) and Liu et al. (1995), who updated the recombination coefficients to account for deviations from LS-coupling. The values for C II are from Davey, Storey, & Kiselius (2000).

Table 4 shows the abundances derived for He I, He II, and C II for each nebula. Table 5 shows the abundances derived for O II for each nebula, ordered by wavelength. A weighted average and weighted standard deviation is calculated for the O II RL abundance for each nebula, using the line fluxes as weights. Lines that were not detected at 3σ significance or better were not included in the average. We note that the small error on the weighted averages reflect the statistical and measurement errors, but does not include errors in the recombination coefficients. It is therefore likely that the actual error would be larger.

In the low-density limit, the ion abundance, relative to hydrogen, from a collisionally excited line can be expressed in the analytic form

$$\frac{N(X^{+n})}{N(H^{+})} = \frac{\lambda(X)}{\lambda(H\beta)} \frac{I(\lambda)}{I(H\beta)} \frac{\alpha_{\text{eff}}(H\beta)}{q_{\text{coll}}(\lambda)} , \tag{5}$$

where $q_{coll}(\lambda)$ is the collisional transition rate, given by

$$q_{coll}(\lambda) = \frac{8.63\times10^{-6}\,\Omega(1,2)}{T^{1/2}\omega_1} e^{-\chi/kT_e}. \tag{6}$$

$\Omega(1,2)$ is the collision strength, $\omega_1$ is the statistical weight of the lower level, and $\chi$ is the threshold energy. In this paper, we calculated abundances using *nebular.ionic* (Shaw & Dufour 1995). Abundances determined from [O III] are shown in Table 6. In addition, previous studies have determined the abundances from ultraviolet spectra for C III] λ1909; these data are shown in Table 6, along with the corresponding references. The errors on the CELs include the errors in electron temperature and density determinations.

## 4. RESULTS

In this section, we present abundance results from RLs for O II and C II, and from CELs for [O III], for our PN sample. We will look at the systematic properties of the difference in abundances from RLs and CELs, and determine how the abundance discrepancy varies, if at all, with the physical properties of PNs. For this purpose we enlarge the PN sample by including published data from the literature, including results from Liu et al. (1995, 2000, 2001), and Tsamis et al. (2003, 2004).

### 4.1. *Distribution of Abundance Discrepancies in O II and C II*

With our new data, we find that there is a considerable spread in the difference in abundances from RLs and CELs. Here we will define the quantity $\Delta(O^{+2})$ = $\log (O^{+2}/H)_{O\,II} - \log(O^{+2}/H)_{[O\,III]}$, with $\Delta(C^{+2})$ defined similarly for carbon. Figure 1 shows a histogram of $\Delta(O^{+2})$ for the full sample of PNs from our study and the published data from Liu et al. (1995), Liu et al. (2000), and Liu et al. (2001), Tsamis et al. (2003), and Tsamis et al. (2004). We see that $\Delta(O^{+2})$ ranges from 0.1 dex to 1.4 dex. The two objects with the smallest abundance differences, NGC 6790 and NGC 6572, are found in our sample. The considerable range in $\Delta(O^{+2})$ suggests that errors in the radiative recombination coefficients for O II are not a cause for the abundance discrepancy, as already noted by Liu et al. (2000).

### 4.2 *Comparison of O II Abundances by Multiplet*

In Figure 2 we show the $O^{+2}$ abundance by emission line for the O II spectrum, relative to the average $O^{+2}$ abundance. The error bar for each point shows the standard deviation of the mean for the sample. Figure 2 shows that the $O^{+2}$ abundances derived from individual O II lines typically fall within $1\sigma$ of the average abundance. This reinforces the conclusion that errors in O II recombination coefficients are not the source of the RL-CEL abundance discrepancy. One significant exception concerns abundances derived from the O II multiplet V15 lines at 4590, 4596 Å. Figure 2 shows that the $O^{+2}$ abundance derived from these lines deviates significantly from the average for other O II lines. A key factor may be that multiplet V15 is populated primarily by dielectronic recombination (Nussbaumer & Storey 1984; Storey 1994). However, it should be noted that we have used the low-temperature dielectronic recombination coefficients for multiplet V15, appropriate for nebular conditions, computed by Nussbaumer & Storey (1984). Storey (1994) did not publish recombination coefficients for V15, but did remark that the computed values were smaller than $10^{-14}$ cm$^3$ s$^{-1}$, consistent with the calculations of Nussbaumer & Storey (1984). Therefore, it is puzzling that the V15 lines should yield such discrepant abundances. Either the recombination coefficients for these transitions are in error, or some physical process is overpopulating the levels that give rise to multiplet V15. It is possible that such a mechanism could overpopulate O II levels in general.

### 4.3 *Correlations with Nebular Properties*

The RL-CEL discrepancy $\Delta(O^{+2})$ shows an inverse correlation with the Balmer surface brightness of the nebula, which is proportional to the emission measure of the nebula. Figure 3 shows $\Delta(O^{+2})$ plotted against the $H\beta$ surface brightness, $S(H\beta)$. The surface brightness was calculated using $H\beta$ fluxes from Cahn et al. (1992), and angular diameters are from Tylenda et al. (2003), Acker et al. (1992), and direct measurement from Hubble Space Telescope images[2], for NGC 6752 , NGC 3132, IC 4406, and NGC 6790. The $H\beta$ fluxes were corrected for reddening using values for $c(H\beta)$ from Cahn et al. (1992). Figure 3a shows data from this paper, Liu et al. (1995), Liu et al. (2000), and Liu et al. (2001), while Figure 3b adds results from Tsamis et al. (2003, 2004). A very tight correlation between $\Delta(O^{+2})$ and the $H\beta$ surface brightness is seen in Figure 3a, although the Tsamis et al. (2003) data in Fig. 3b show larger scatter. Least-squares linear fits to the data in Figures 3a and 3b are shown. For Figure 3a, the fit is given by

$$\Delta(O^{+2}) = -(0.386 \pm 0.036)\log S(H\beta) + (1.38 \pm 0.06)\,, \qquad (7)$$

with a correlation coefficient of -0.82; for Figure 3b, the fit is given by

$$\Delta(O^{+2}) = -(0.408 \pm 0.075)\log S(H\beta) + (1.26 \pm 0.15)\,, \qquad (8)$$

---

[2] http://heritage.stsci.edu/index.html

with a correlation coefficient of -0.60. The trend with surface brightness is mirrored in a direct correlation between $\Delta(O^{+2})$ and linear diameter (Garnett & Dinerstein 2001), although this correlation is not as tight because of the greater uncertainty in distances and thus linear diameters. Figure 4 shows the same plot as Figure 3 for $C^{+2}$, using abundances derived from C II λ4267 and C III] λ1909. A similar trend of $\Delta(C^{+2})$ with $H\beta$ surface brightness is seen. A linear least-squares fit is shown, given by

$$\Delta(C^{+2}) = -(0.431 \pm 0.072)\log S(H\beta) + (1.30 \pm 0.14)\,, \tag{9}$$

with a correlation coefficient of -0.71. The increased scatter is not unexpected, because whereas O II and [O III] can be observed in a single spectrograph setting, sampling identical areas of a nebula, C II and C III] are found in very different spectral regions; C II is in the visible spectrum, while C III] is in the UV. Most C III] measurements were obtained with *IUE*, which has a much larger aperture than typical visible-light spectrographs, so the areas subtended by C II and C III] observations can be very different.

The trends seen in Figures 3 and 4 are very similar for both oxygen and carbon; compact, high surface brightness nebulae have smaller RL-CEL abundances discrepancies than extended, faint nebulae. This correlation suggests that nebular density might be an important factor, since the Balmer line surface brightness is proportional to $n_e^2 dl$. Figure 5 shows that the abundance discrepancy is inversely correlated with electron density as derived from forbidden-line ratios. A linear least-squares fit is shown, given by

$$\Delta(O^{+2}) = -(0.635 \pm 0.134)\log n_e + (2.89 \pm 0.51)\,, \tag{10}$$

with a correlation coefficient of -0.47. In Figure 6, we plot the abundance discrepancy versus diameter, where we have derived linear sizes for the sample PNs based on statistical distances derived using the methods of Zhang (1995), Van de Steene & Zijlstra (1995) and Bensby & Lundström (2001). These recent studies use improved sets of distance calibrators to derive statistical relations between angular size and radio flux density, and appear to provide much better distances than older studies (see Phillips (2002) for an excellent comparison of various PN distance calibrations; however, note that the distance scale derived by Phillips appears to be systematically smaller by a factor 2.7 from all of the comparison studies). Based on these distances, Figure 6 shows an excellent correlation between nebular diameter $D$ and the $O^{+2}$ abundance discrepancy. The linear least-squares fit is given by

$$\Delta(O^{+2}) = (3.15 \pm 0.56)D + (0.035 \pm 0.098)\,, \qquad (11)$$

with a correlation coefficient of 0.67. The strength of this correlation contrasts with Tsamis et al. (2004), who found only a weak correlation. We suspect that the weaker correlation they see is the result of their use of the distance calibration of Cahn, Kaler, & Stanghellini (1992), which is similar to those of Van de Steene & Zijlstra (1995) and Bensby & Lundström (2001), but has greater scatter due to the use of older calibration data.

In contrast to the above results in Figure 3, Figure 7 shows that the $O^{+2}$ abundance derived from O II multiplet V15 deviates strongly from the average O II abundance for the high surface brightness nebulae, but this discrepancy decreases and ultimately disappears as the surface brightness decreases. Figure 7a shows data from this paper, Liu et al. (1995), Liu et al. (2000), and Liu et al. (2001), while Figure 7b adds results from Tsamis et al. (2003, 2004). A very tight correlation between Δ(V15) and H$\beta$ surface brightness is seen in Figure 7a, although the Tsamis et al. (2003, 2004) data in Fig. 7b show larger scatter. However, the three points of low surface brightness from Tsamis et al. (2003) that fall outside of the trend outlined in Figure 7a had only one measurement of the two observable lines of multiplet V15. Since the 4590Å and 4596Å lines of multiplet V15 have comparable intensities, the detection of only one line suggests that the measurement may be only an upper limit, and thus may not give a reliable estimate of the multiplet V15 intensities in those cases where only one line is measured. Comparison of Figures 3 and 7 raises two questions: (1) Why does multiplet V15 behave differently from the other O II lines? (2) Why does the $O^{+2}$ abundance for V15 differ so much from the average for other O II lines, despite the fact that we have used the appropriate rate for dielectronic recombination from Nussbaumer & Storey (1984)? We ask whether there might be more than one contribution accounting for the different behavior of the V15 lines. We start by looking at the highest surface brightness nebulae, NGC 6572 and NGC 6790, where we note that the [O III] and O II abundances are in relatively good agreement (excluding the V15 lines). We then make the assumption that the recombination coefficient for multiplet V15 is in error, and estimate a new value which forces the V15 abundances into agreement with the average O II abundance for NGC 6572 and NGC 6790. Next, we take this new value for the V15 recombination coefficient and estimate for each of the other PNs the V15 line strength that would be predicted based on the abundance derived from [O III], subtract that from the observed V15 line strength, and compute the $O^{+2}$ abundance from the residual V15 line. The result is shown in Figure 8, which shows the difference between the [O III] abundances and the residuals of the V15 multiplet abundances after taking out the contribution described above. Comparing Figures 8 and 3, we see that plotting the residual V15 abundance vs. $H\beta$ surface brightness recovers the trend seen for the other O II lines in Figure 3, although with larger scatter. Even if the three discrepant points of Figure 7b from Tsamis et al. (2003) are removed, the trend of Figure 3 is still recovered. This implies that there are two contributions to the V15 multiplet, one that has a rate that is roughly constant for all nebulae, and one which follows the same behavior as the other O II lines. Based on this result, we suggest the possibility that the dielectronic recombination rates for V15 are too small by a factor 4-5, either because the coefficients are in error or because there is a physical process not accounted for that increases the recombination rates. It is likely that the coefficients from Nussbaumer and Storey (1984) are in error, as noted by Liu et al. (2001).

A new study of dielectronic recombination rates by Zatsarinny et al. (2004) sheds some light on the uncertainties in the rates. They have computed new dielectronic recombination rates for ions in the carbon sequence, using a multi-configuration Breit-Pauli approach under intermediate coupling, and have kindly made the rates available on a web site (http://www-cfadc.phy.ornl.gov/data_and_codes). We have compared the new computed rates for multiplet V15 with those of Nussbaumer & Storey (1984), which were computed under the assumption of LS coupling. The branching ratio for V15 was determined using transition probabilities from the NIST atomic database. Assuming that the recombining $O^{+2}$ ions are all in the ground $^3P_0$ state, we find that the new $\alpha_{eff}$(DR) for multiplet V15 from Zatsarinny et al. (2004) is a factor 1.8 larger than the rate from Nussbaumer & Storey (1984) at both 8,000 K and 20,000 K. This larger DR rate for multiplet V15 reduces the derived abundances, but does not bring the abundances from the V15 lines into agreement with the average for the other O II lines in our study. Other multiplets in the O II recombination spectrum are affected by dielectronic recombination also, and it is of interest to determine the impact of the new DR rates on other lines. For O II multiplet V1, for example, we find that the new DR rate at T = 10,000 K is about $8\times10^{-15}$ $cm^3$ $s^{-1}$, which is several times larger than the rate from Nussbaumer & Storey (1984). However, this new DR rate is only about 2% of the radiative recombination rate for multiplet V1 from Storey (1994), so the impact of the new DR calculation on the observed V1 emission is negligible. A complete comparison of rates for all of the lines in the O II spectrum is beyond the scope of this paper, but based on these examples we expect that the effect of the new DR rates will be small for other lines populated mainly by radiative recombination.

Of additional importance is that the new total DR rates for $O^{+2}$ of Zatsarinny et al. (2004) are much larger at low temperatures than their values under LS coupling, which they attribute to the inclusion of relativistic effects. This effect grows as the temperature decreases, so that at T = 100 K the new DR rate exceeds the LS value by approximately a factor ten, and even exceeds the radiative rate by a factor three. The Zatsarinny et al. DR rate becomes comparable to the radiative rate at a temperature of about 7,000 K. This could have a significant effect on the interpretation of the O II spectrum, and so some reevaluation of the recombination line abundances may need to be made.

Mechanical energy deposition by shocks is a possible mechanism for increasing the heating in a nebula and thus the excitation of forbidden lines. One manifestation of shocks is high-velocity material. A rapidly-expanding nebula will experience stronger shocks as it plows into the ambient interstellar medium. Figure 9 plots the $O^{+2}$ discrepancy versus expansion velocity obtained from Weinberger (1989) and Gesicki and Zijlstra (2000). For this study, we used the [O III] expansion velocities. Figure 9 shows that the abundance discrepancy shows a weak correlation with expansion velocity, such that the discrepancy is greater in nebulae with the fastest expansion rates. However, shock-heated gas should also show higher electron temperatures. Figure 10 plots the $O^{+2}$ abundance discrepancy versus electron temperature $T_e$, as derived from [O III]. We see no apparent correlation of the RL-CEL abundance discrepancy with electron temperature.

On the other hand, Liu et al. (2001) demonstrated a close correlation between the temperature difference T[O III] – T(Balmer jump) and the RL-CEL abundance discrepancy. The origin of this temperature difference is not yet understood; it is not necessarily expected that the Balmer temperature should equal the [O III] temperature, but the observed differences are in some cases much larger than can be accounted for by photoionization and thermal equilibrium. Stasinska & Szczerba (2001) have pointed out that photoelectric heating by dust grains in the relatively hard UV radiation field near a PN central star can boost $T_e$ in the center of the nebula; grains have a lower ionization energy than H, so the mean photoelectron energy in a dusty nebula is higher than in a dust-free nebula. This should be a fruitful area for future investigation.

We see no relation between the RL-CEL abundance discrepancy and the effective temperature of the central star (Figure 11). Temperatures are from Kaler & Jacoby (1991), Preite-Martinez & Pottasch (1983), Preite-Martinez et al. (1989), and Preite-Martinez (1993), using an average of results generated by an inhomogeneous set of methods, including energy balance, continuum fitting, and Zanstra methods. Nor does the discrepancy correlate with nebular ionization (or “excitation class”); in Figure 12 we plot the abundance discrepancies against the $He^{+2}/He^{+}$ ratio derived from the optical spectra. This plot shows a poor correlation between the $O^{+2}$ abundance discrepancy and the He ionization. This disfavors exotic processes such as charge exchange between $He^{+}$ and $O^{+2}$ in a hot central bubble as a mechanism to enhance the O II recombination rate in the nebular interior; this charge exchange process is highly endothermic at nebular temperatures. Finally, we find no relationship between the abundance discrepancy and the spectral type of the central star; large abundance discrepancies are seen in both O and WR type central stars. Thus, mass loss rate and stellar surface abundances appear not to play a role in the RL-CEL discrepancy.

We have looked at morphological class to see if the RL-CEL abundance discrepancy is related to the progenitor. For instance, Type I planetary nebulae, those with high N and He abundances, are strongly correlated with bipolarity (Torres-Peimbert & Peimbert 1997), and are believed to have more massive AGB progenitors with perhaps higher mass loss rates and faster winds. Meanwhile, Stanghellini et al. (2002) have shown evidence that elliptical (E) and round (R) PNs have a larger scale height above the disk than bipolar (BP) PNs, also suggesting that the E and R types are associated with older, less massive progenitors than the BP types. Figure 13 shows the distribution of the RL-CEL abundance difference for PNs of the different classes.

The PN morphologies were taken from Stanghellini et al. (2002) for PNs in common with their sample, otherwise we classified the nebula on the basis of images from Balick (1987), or *Hubble Space Telescope* images from the STScI EPO site. We combined E and R types together to improve the number statistics. Figure 13 shows that there is no noticeable difference in the distribution of $O^{+2}$ abundance differences for the morphological types; the mean and standard deviation for E+R types is the same as that for BP PNs, within the errors.

*4.4 Relation Between $O^{+2}$ and $C^{+2}$ Abundance Discrepancies*

Liu et al. (2000) argued that the RL-CEL discrepancies had essentially the same value for $O^{+2}$, $C^{+2}$, $N^{+2}$, and $Ne^{+2}$ in a given nebula. In other words, the magnitude of the abundance discrepancy was the similar regardless of which element is observed. However, this claim was based on only two observed PNs at the time. Garnett & Dinerstein (2001) pointed out that NGC 6720 showed different abundance discrepancies for $O^{+2}$ and $C^{+2}$, and that O II and C II showed different spatial variations. Still, the sample of PNs studied at the time was small, and it is difficult to compare optical narrow-slit C II measurements and IUE-based UV measurements.

We re-examine this question using the larger sample of PNs compiled here. Figure 14 shows a comparison of the $O^{+2}$ and $C^{+2}$ abundance discrepancy for all 22 PNs examined to date. The line in the figure shows equality. This plot shows that most of the PNs in the sample follow the line of equality to within the errors, although two objects, NGC 6302 and NGC 7009, appear to be more than 2$\sigma$ from the line. This result favors the argument by Liu et al. (2000) that the C II and O II abundances are enhanced by the same factor, although the comparison of abundances from C II and C III] are more uncertain because of differences in areas sampled by optical and UV spectrograph apertures. A more extensive comparison of abundances from other species would help resolve this question.

# 5. DISCUSSION

In this paper we have provided data on recombination lines for a new sample of planetary nebulae, and looked for correlations between the RL-CEL abundance discrepancy and various nebular properties, in the hope that others can use the results to provide a physical explanation for the abundance discrepancy. In the end, only a few notable correlations are seen. The first is the correlation between the abundance discrepancy and nebular diameter/surface brightness (Figures 3 and 4). The second is the correlation between the abundance discrepancy and the temperature difference T[O III] – T(Balmer jump), pointed out in Liu et al. (2001) and Tsamis et al. (2004). A third correlation is seen between the abundance discrepancy and the electron density. A correlation seen here between the abundance discrepancy and expansion velocity is of marginal significance.

Garnett & Dinerstein (2001), and Liu et al. (2000) have presented spatially-resolved measurements of RLs and CELs in NGC 6720 and NGC 6153, and found that the ratio of the abundances derived from RLs and CELs varies across both nebulae. The largest differences, curiously, are found towards the center of the nebula in each case. This seems to rule out temperature fluctuations as a source of the discrepancy, as the temperature would have to be lower in the center of the nebula, contrary to what is expected in general from PN ionization models. In addition, while it would be expected that the O II RL emission should peak in the transition between the $O^{+}$ and $O^{+2}$ zones, the spatially-resolved studies done so far suggest that the RL emission actually peaks in the highly-ionized inner regions of PNs. However, since only two nebulae have well-resolved measurements of recombination lines, it is not possible to generalize these results. The spatial distribution of forbidden line and recombination line emission offers clues to the physical processes that dominate the emission, so studies of more objects at high spatial resolution are needed as well as high spectral resolution data.

Any model to explain the RL-FL abundance discrepancy needs to account for not only the absolute value of the discrepancy but also for the correlations with nebular properties and for the spatial variations seen within individual nebulae. For example, Liu et al. (2000) proposed that cold, dense, H-poor knots embedded within the ionized gas could account for enhanced recombination line emission. One can ask, however, why the evidence for these knots only appears in larger, more evolved nebulae. Also, in NGC 6720, why are the largest RL enhancements observed to be not coincident with the dust knots seen in HST imaging, but rather in the highly ionized region where $He^{+2}$ dominates (Garnett & Dinerstein 2001)? On the other hand, can a model with enhanced dielectronic recombination (Garnett & Dinerstein 2001) account for enhanced recombination lines from so many ions? What is the origin of the sometimes large difference between the temperature derived from [O III] and that derived from the Balmer jump? Finally, do temperature fluctuations (Peimbert 1967) play much of a role, if at all, in the RL-CEL discrepancy? This appears to be increasingly unlikely. The work of Liu et al. (2000) and Tsamis et al. (2003, 2004), comparing collisionally-excited lines in optical, UV, and IR spectra, have demonstrated that the abundances derived from these different transitions generally do not differ greatly, despite the large differences in excitation energies, and that the differences do not correlate at all with the RL-CEL abundance difference or with the difference in excitation energies for the transitions. This result conflicts with the expected behavior for forbidden line abundances in the case of temperature fluctuations, and argues strongly against them as significant factor in the RL-CEL discrepancy.

We also suggest that there may still be room for improvement in the recombination rate coefficients for elements heavier than H and He. A recent study by Sharpee, Baldwin, & Williams (2004) compared C II, N II, and O II line strengths for a number of PNs. They noted that there do appear to be significant variations in the strengths of recombination lines above what might be expected from purely observational scatter, and they suggested that additional physical processes beyond recombination might be responsible. We have discussed here the new DR rate coefficients from Zatsarinny et al. (2004). We encourage continued work to improve both radiative and dielectronic recombination rates for the important ions observed in ionized nebulae. New observations and atomic rate calculations are both needed to resolve the large, puzzling discrepancy between recombination line and forbidden line abundances.

We thank Daniel Savin for alerting us to the new work on DR rate coefficients. We thank our referee for pointing out several suggestions that improved the presentation of the paper. This study made use of the NIST Atomic Spectra Databases. Support from NSF grant AST-0203905 and HST grant GO-09839.01-A is gratefully acknowledged.

FIG. 1. – A histogram of the $\Delta O^{+2}$ discrepancy, defined by log $(O^{+2}/H)_{O\ II}$ – log$(O^{+2}/H)_{[O\ III]}$, for the planetary nebulae from our sample plus those of Liu et al. (1995, 2000, 2001) and Tsamis et al. (2003, 2004).

FIG. 2. – The difference between the abundance derived from individual O II lines and the average $O^{+2}$ abundance derived from all O II recombination lines except multiplet V15. Multiplet V15 (λ4590-96) is seen to give significantly higher abundances.

FIG. 3. – The $\Delta O^{+2}$ discrepancy plotted against the $H\beta$ surface brightness. High surface brightness nebulae show the smallest discrepancy. Figure 3a shows the data from this paper and Liu et al (1995, 2000, 2001). Figure 3b adds the data from Tsamis et al. (2003, 2004), which appear to have more scatter.

FIG. 4. – The $\Delta C^{+2}$ discrepancy plotted against the $H\beta$ surface brightness. The trend is similar to that in figure 3.

FIG. 5. – The $\Delta O^{+2}$ discrepancy plotted against the nebular electron density. An inverse correlation is seen.

FIG. 6. – The $\Delta O^{+2}$ discrepancy plotted against the nebular diameter. Larger nebulae have the greatest discrepancies.

FIG. 7. – The discrepancy between the abundance derived from multiplet V15 and the average O II recombination line abundance, plotted versus $H\beta$ surface brightness. Figure 7a shows the data from this paper and Liu et al (1995, 2000, 2001). Figure 7b adds the data from Tsamis et al. (2003, 2004), which appear to have more scatter.

FIG. 8. – The residuals of multiplet V15 versus $H\beta$ surface brightness (see text for description). The trend is similar to that of figure 3.

FIG. 9. – The $\Delta O^{+2}$ discrepancy plotted against the [O III] expansion velocity. The nebulae with greater expansion velocity tend to show greater discrepancy.

FIG. 10. – The $\Delta O^{+2}$ discrepancy plotted against the nebular electron temperature. No correlation is seen.

FIG. 11. – The $\Delta O^{+2}$ discrepancy plotted against the stellar effective temperature. No correlation is seen.

FIG. 12. – The $\Delta O^{+2}$ discrepancy plotted against $He^{+2}/He^{+}$. No correlation is seen.

FIG. 13. – The $\Delta O^{+2}$ discrepancy plotted against the morphological class as obtained from Stanghellini et al. (2002) and Hubble Space Telescope images. There is no apparent distribution in $\Delta O^{+2}$ between the different morphological classes.

FIG. 14. – The $\Delta O^{+2}$ discrepancy plotted against the $\Delta C^{+2}$ discrepancy. The line shows equality. The discrepancies are strongly correlated, suggesting that both the oxygen and carbon abundance discrepancies are driven by the same process.

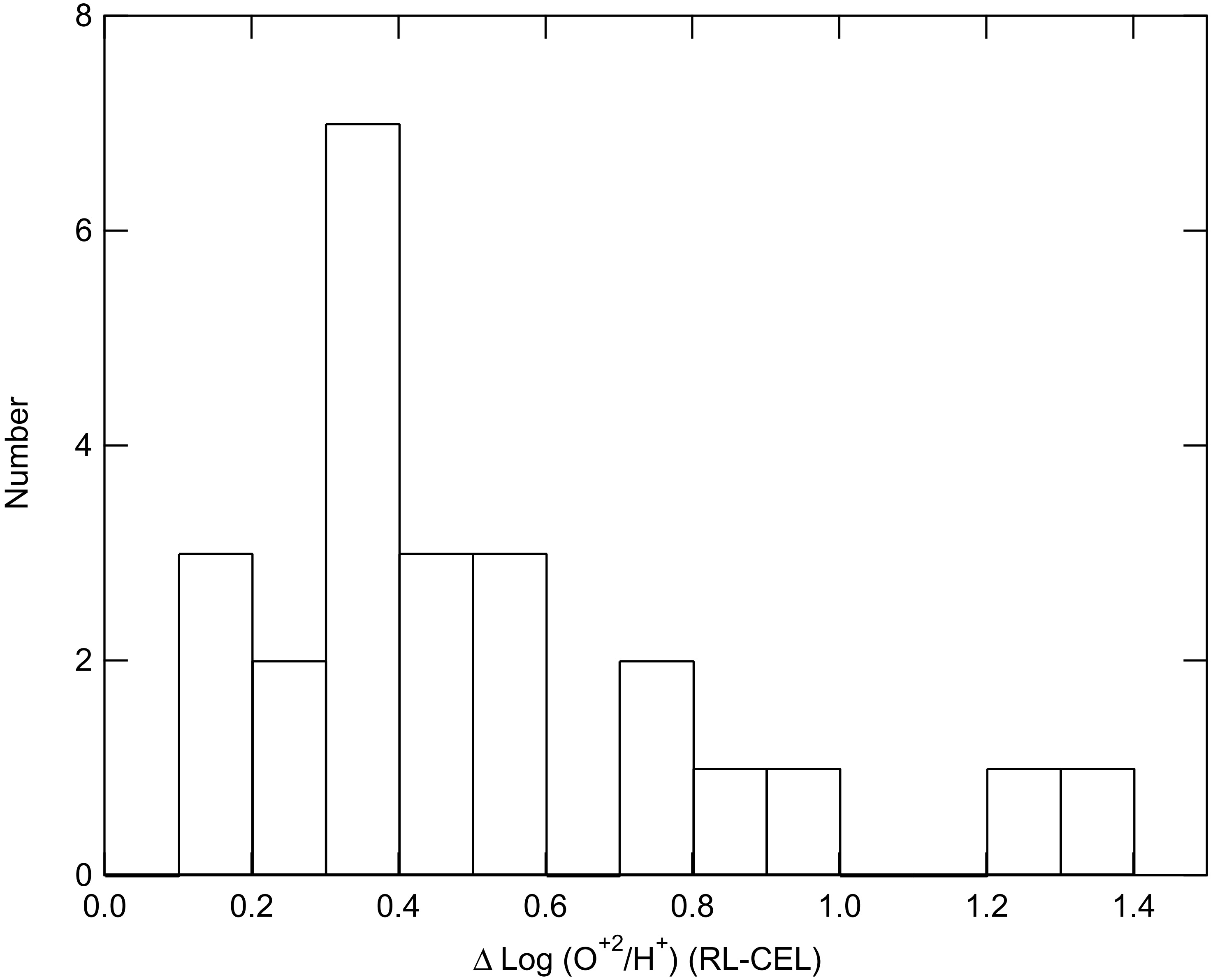

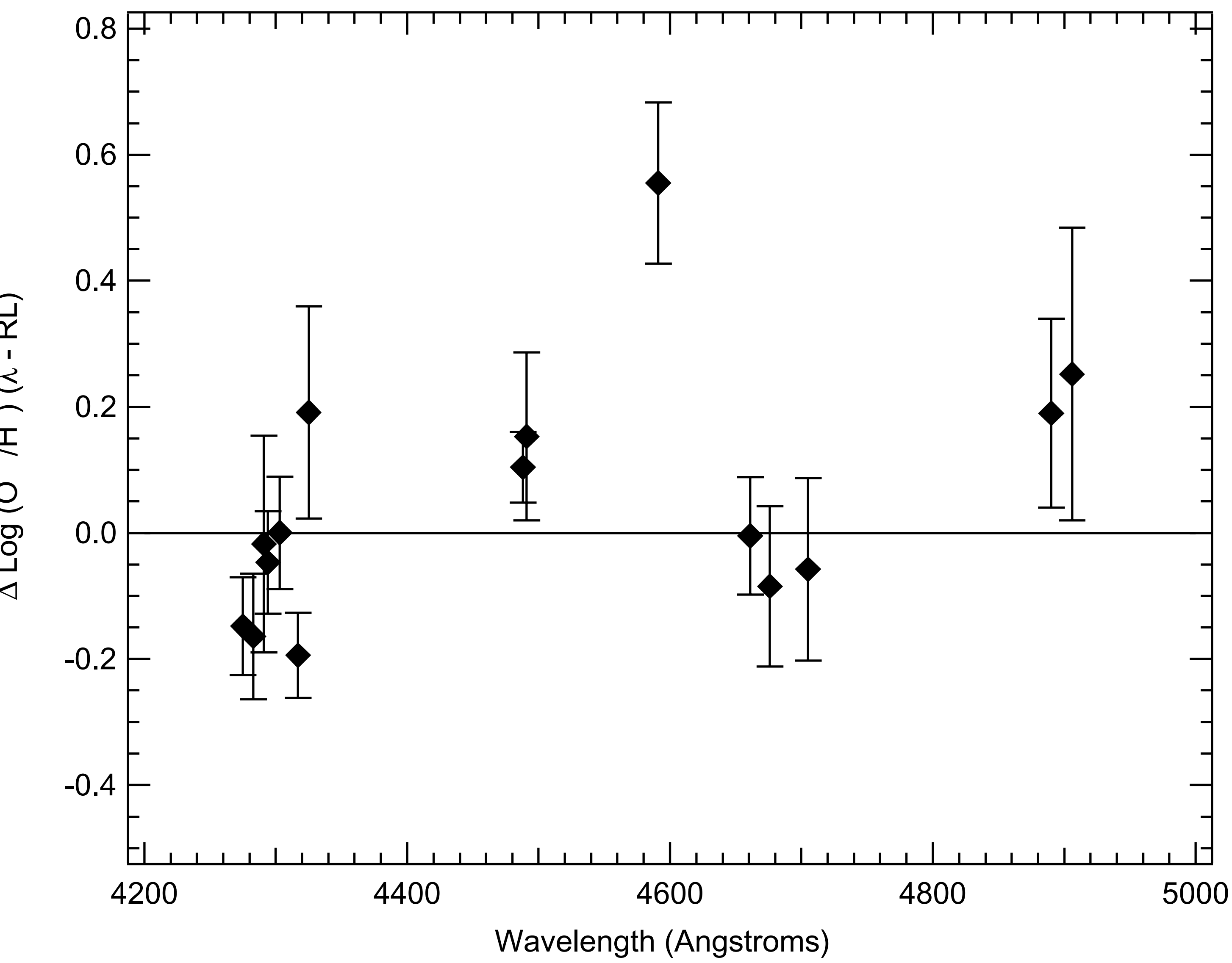

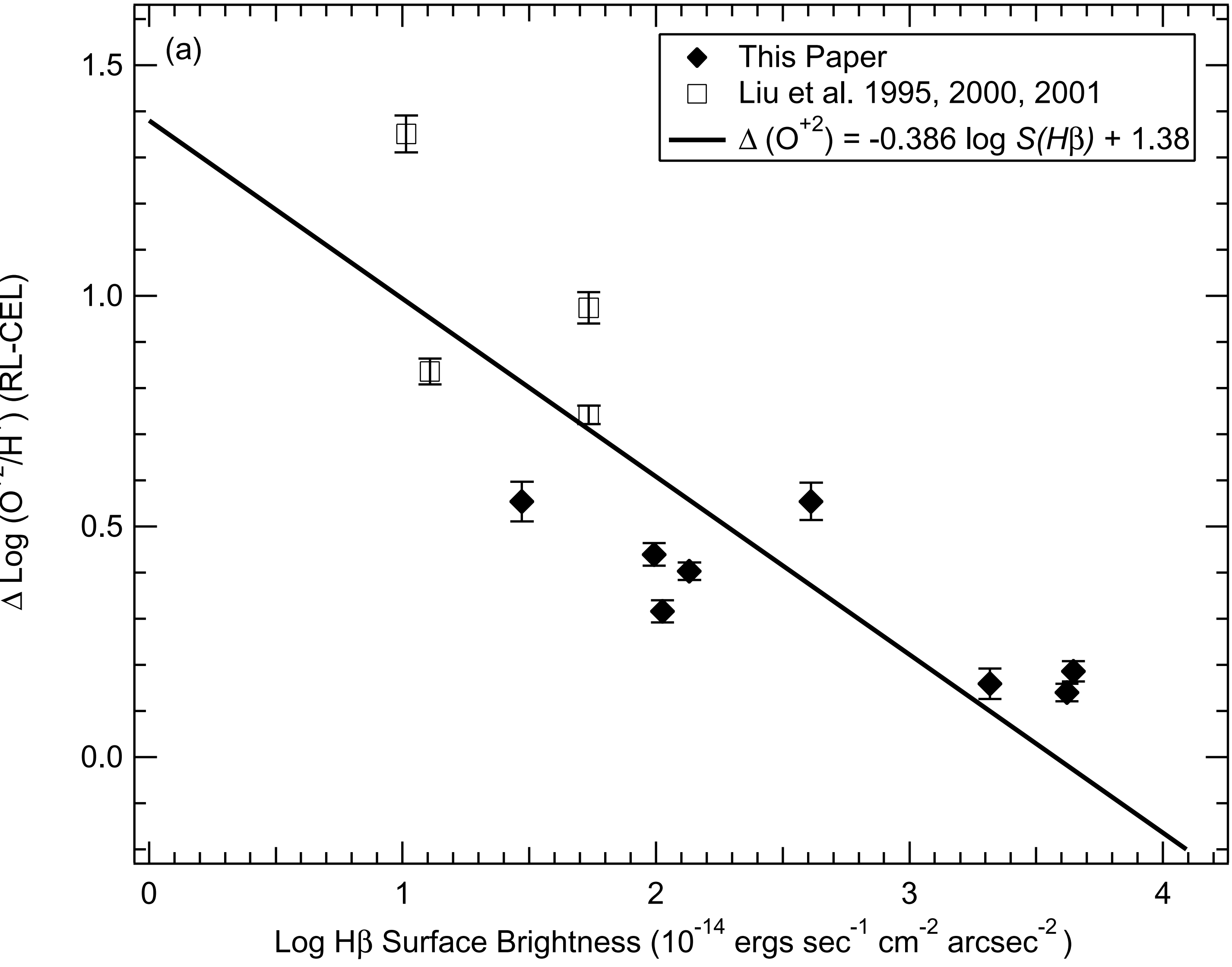

(a)
This Paper
Liu et al. 1995, 2000, 2001
Δ (O+2) = -0.386 log S(Hβ) + 1.38
Δ Log (O+2/H+) (RL-CEL)
Log Hβ Surface Brightness (10-14 ergs sec-1 cm-2 arcsec-2 )
1.5
1.0
0.5
0.0
0
1
2
3
4

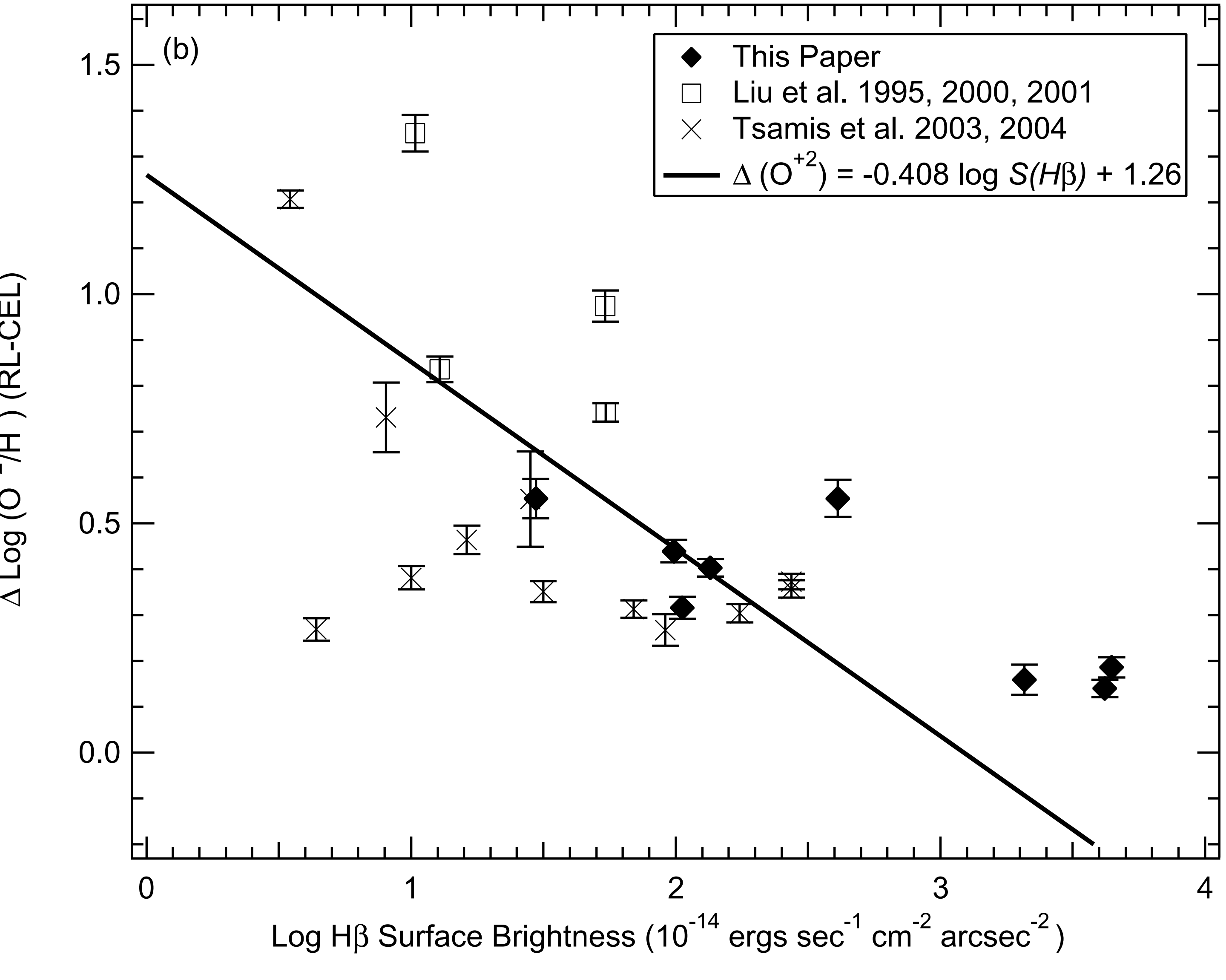

(b)
This Paper
Liu et al. 1995, 2000, 2001
Tsamis et al. 2003, 2004
Δ (O$^{+2}$) = -0.408 log $S(H\beta)$ + 1.26
Δ Log (O$^{+2}$/H$^{+}$) (RL-CEL)
Log Hβ Surface Brightness (10$^{-14}$ ergs sec$^{-1}$ cm$^{-2}$ arcsec$^{-2}$ )
0.0
0.5
1.0
1.5
0
1
2
3
4

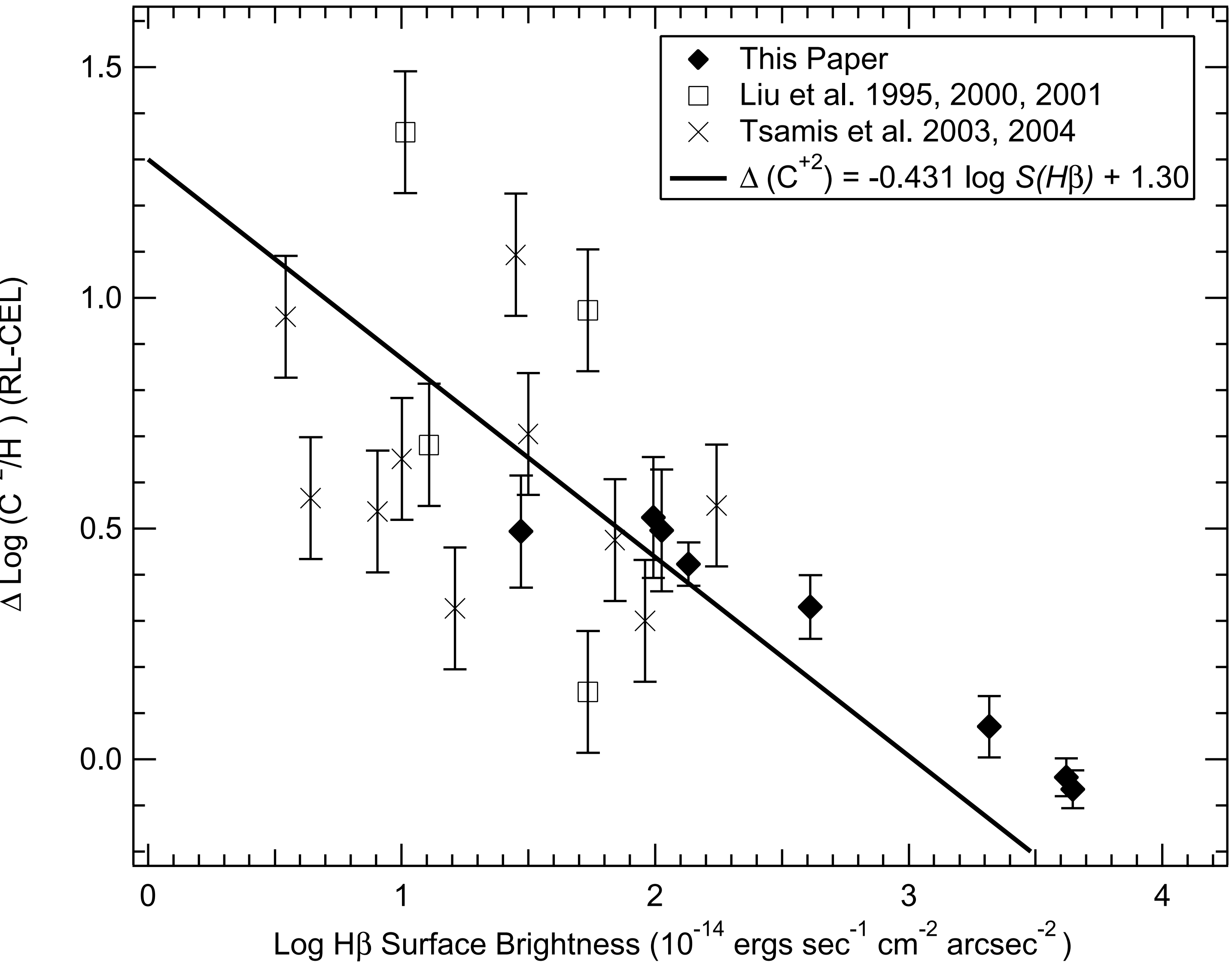

This Paper
Liu et al. 1995, 2000, 2001
Tsamis et al. 2003, 2004
Δ (C+2) = -0.431 log S(Hβ) + 1.30
1.5
1.0
0.5
0.0
0
1
2
3
4
Δ Log (C+2/H+) (RL-CEL)
Log Hβ Surface Brightness (10-14 ergs sec-1 cm-2 arcsec-2 )

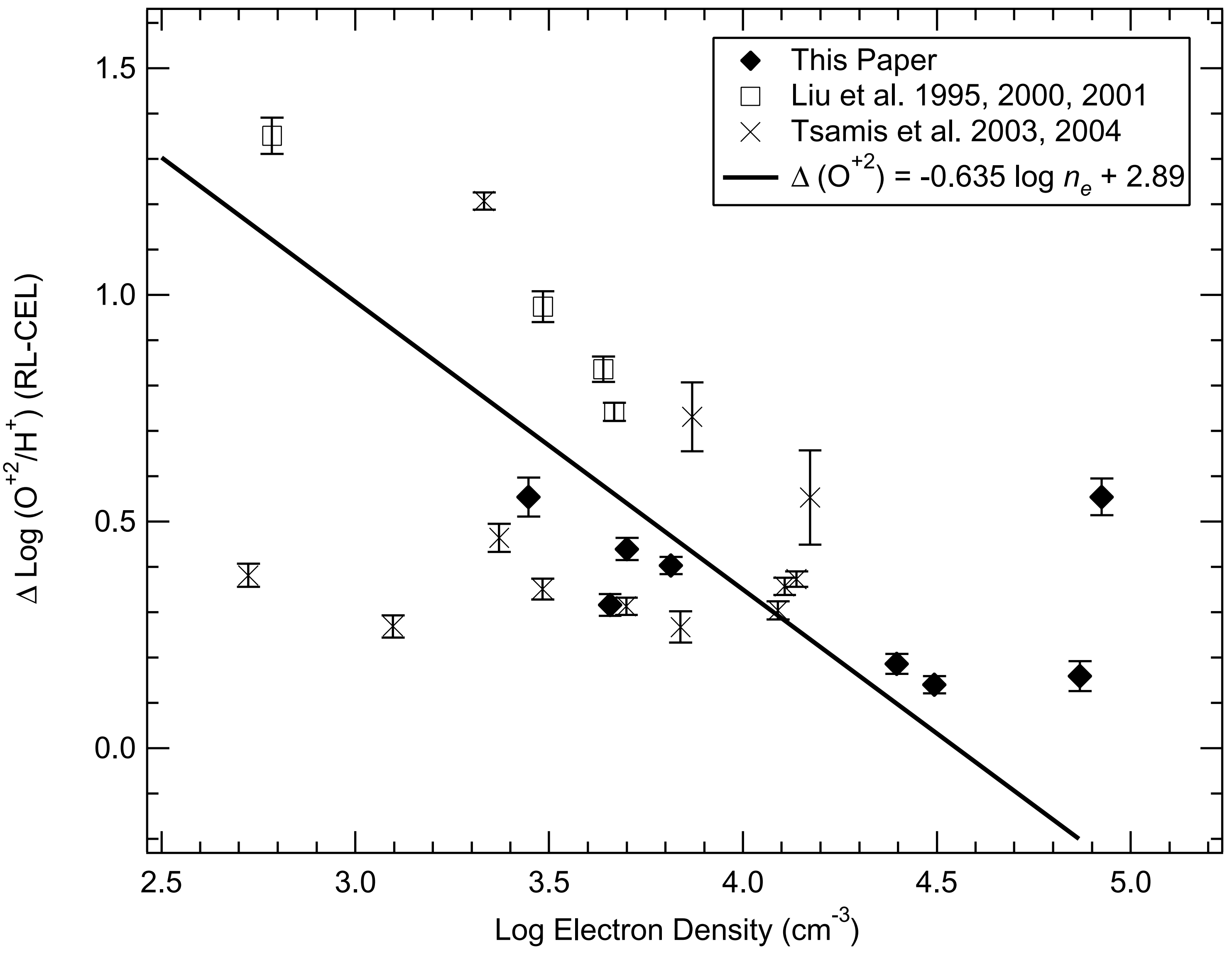

This Paper
Liu et al. 1995, 2000, 2001
Tsamis et al. 2003, 2004
Δ (O+2) = -0.635 log ne + 2.89
Δ Log (O+2/H+) (RL-CEL)
Log Electron Density (cm-3)
1.5
1.0
0.5
0.0
2.5
3.0
3.5
4.0
4.5
5.0

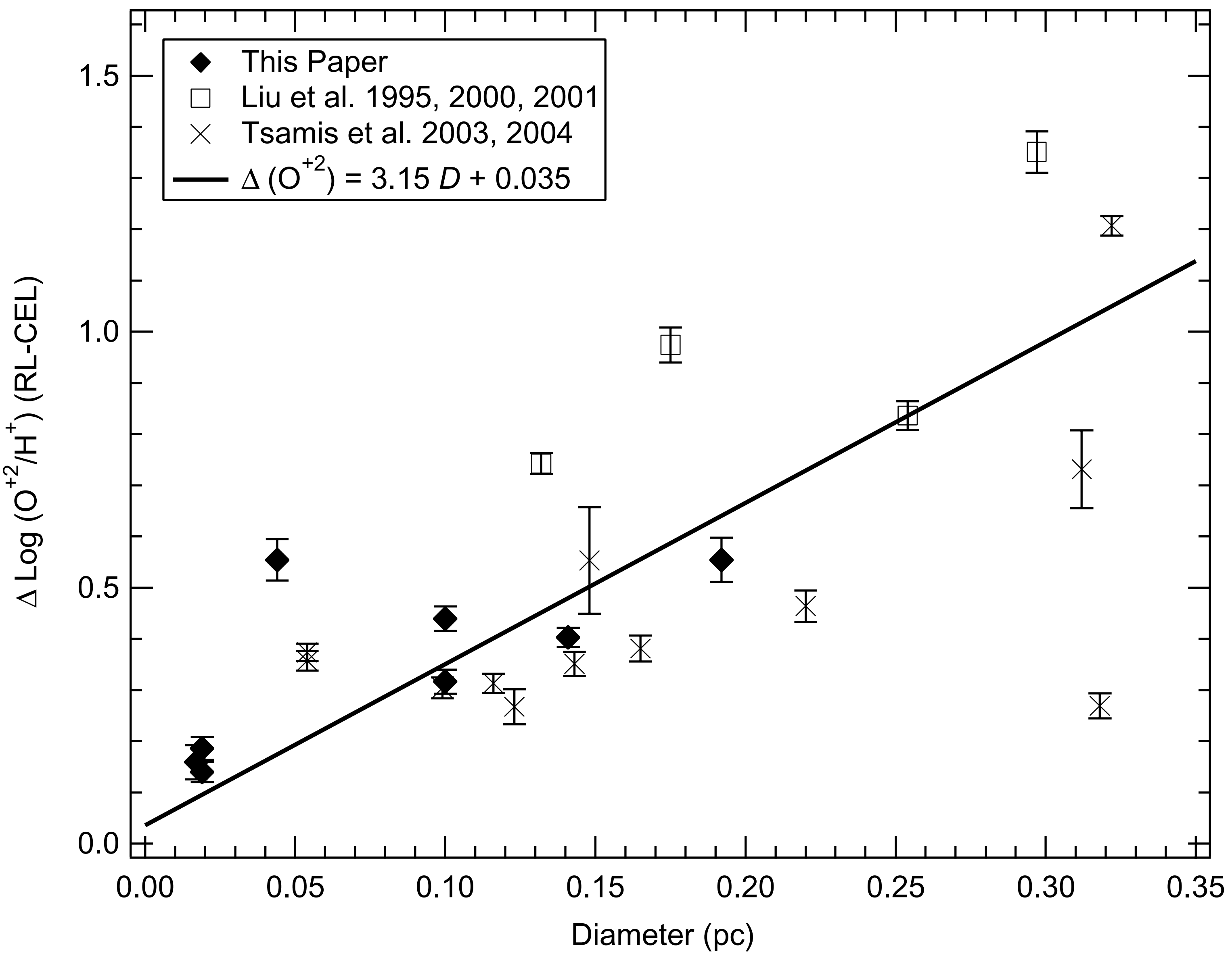

This Paper
Liu et al. 1995, 2000, 2001
Tsamis et al. 2003, 2004
Δ (O+2) = 3.15 D + 0.035
Δ Log (O+2/H+) (RL-CEL)
Diameter (pc)
0.0
0.5
1.0
1.5
0.00
0.05
0.10
0.15
0.20
0.25
0.30
0.35

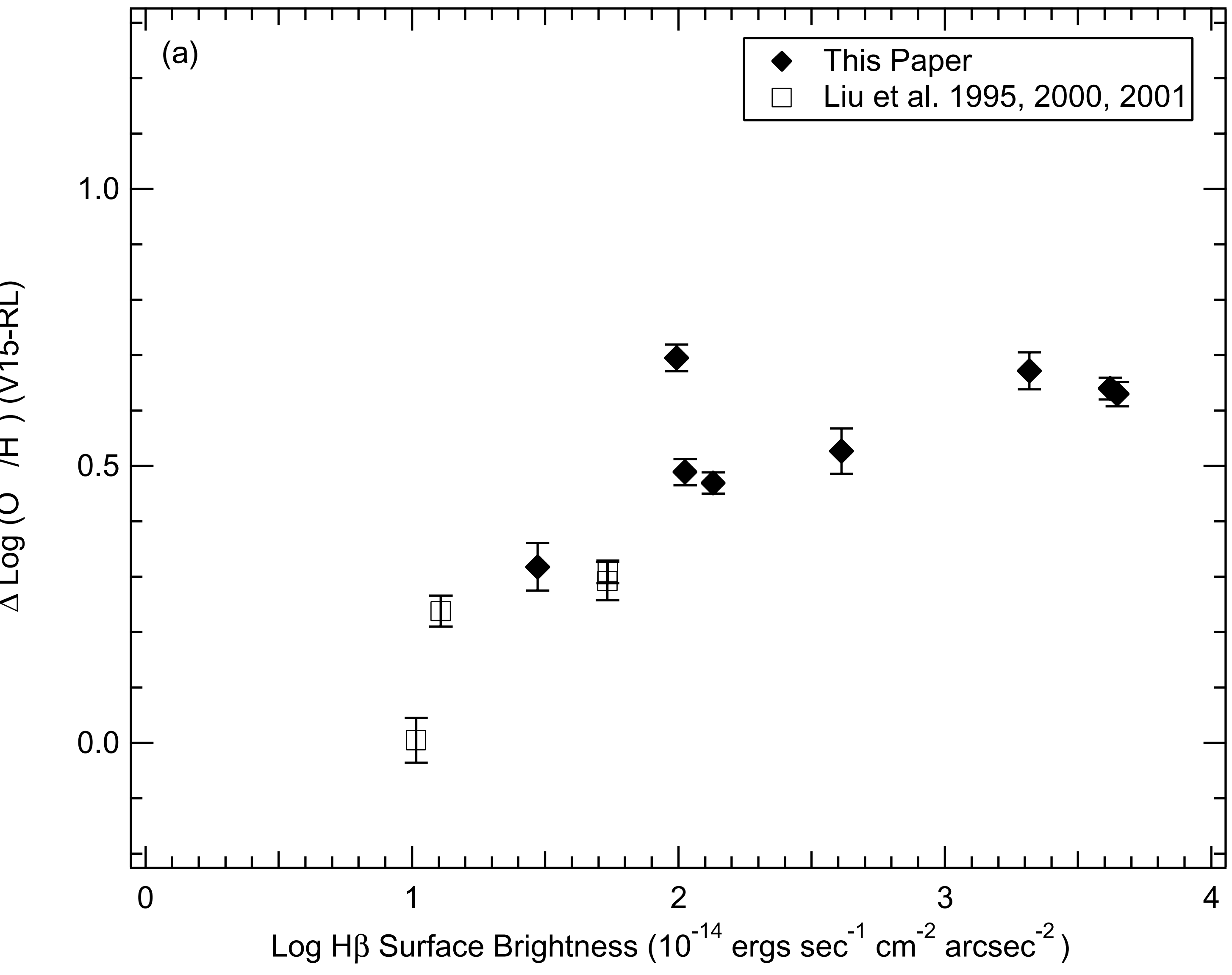
(a)
This Paper
Liu et al. 1995, 2000, 2001
1.0
0.5
0.0
Δ Log (O+2/H+) (V15-RL)
0
1
2
3
4
Log Hβ Surface Brightness (10-14 ergs sec-1 cm-2 arcsec-2 )

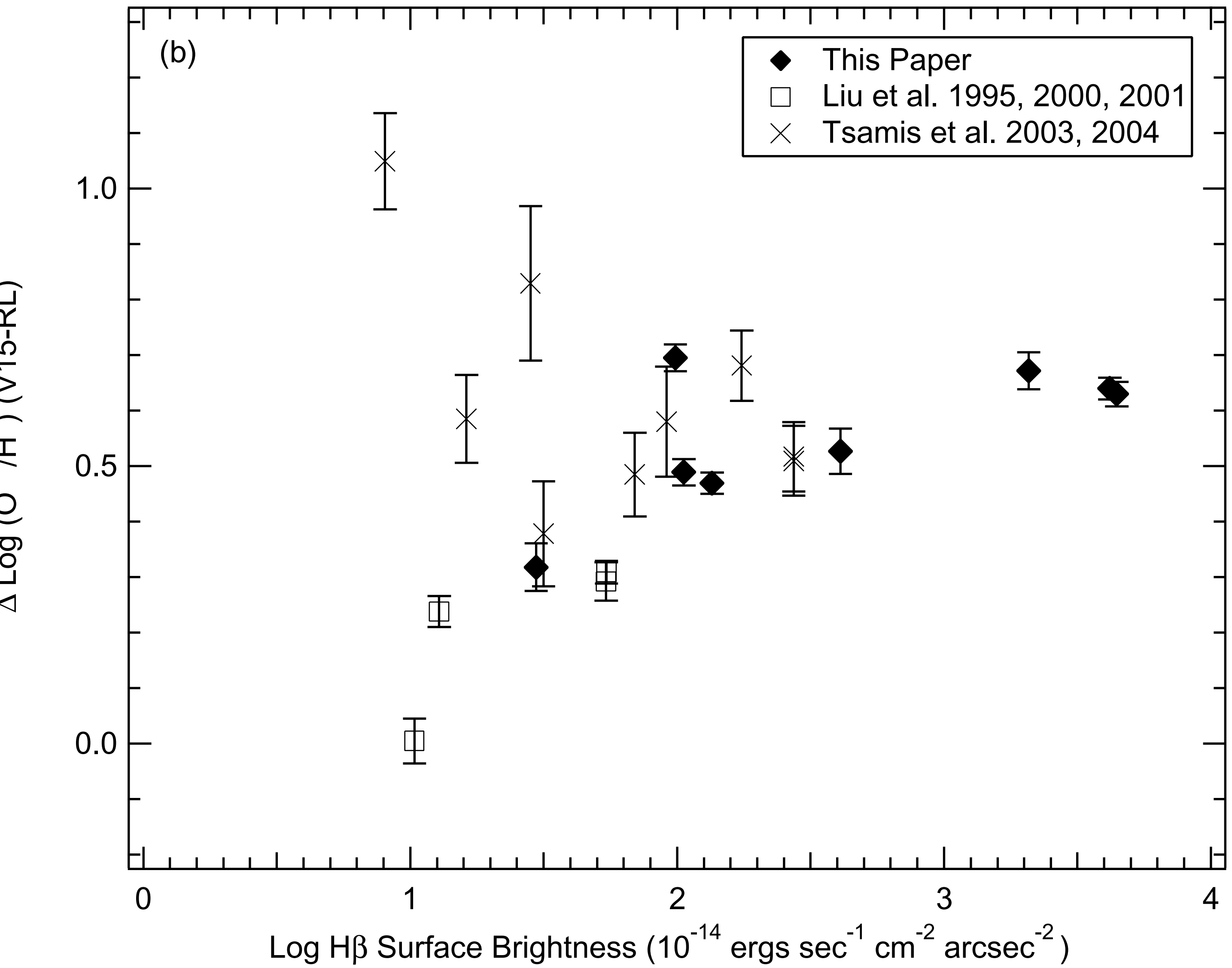

(b)
This Paper
Liu et al. 1995, 2000, 2001
Tsamis et al. 2003, 2004
Δ Log (O$^{+2}$/H$^{+}$) (V15-RL)
1.0
0.5
0.0
0
1
2
3
4
Log Hβ Surface Brightness (10$^{-14}$ ergs sec$^{-1}$ cm$^{-2}$ arcsec$^{-2}$)

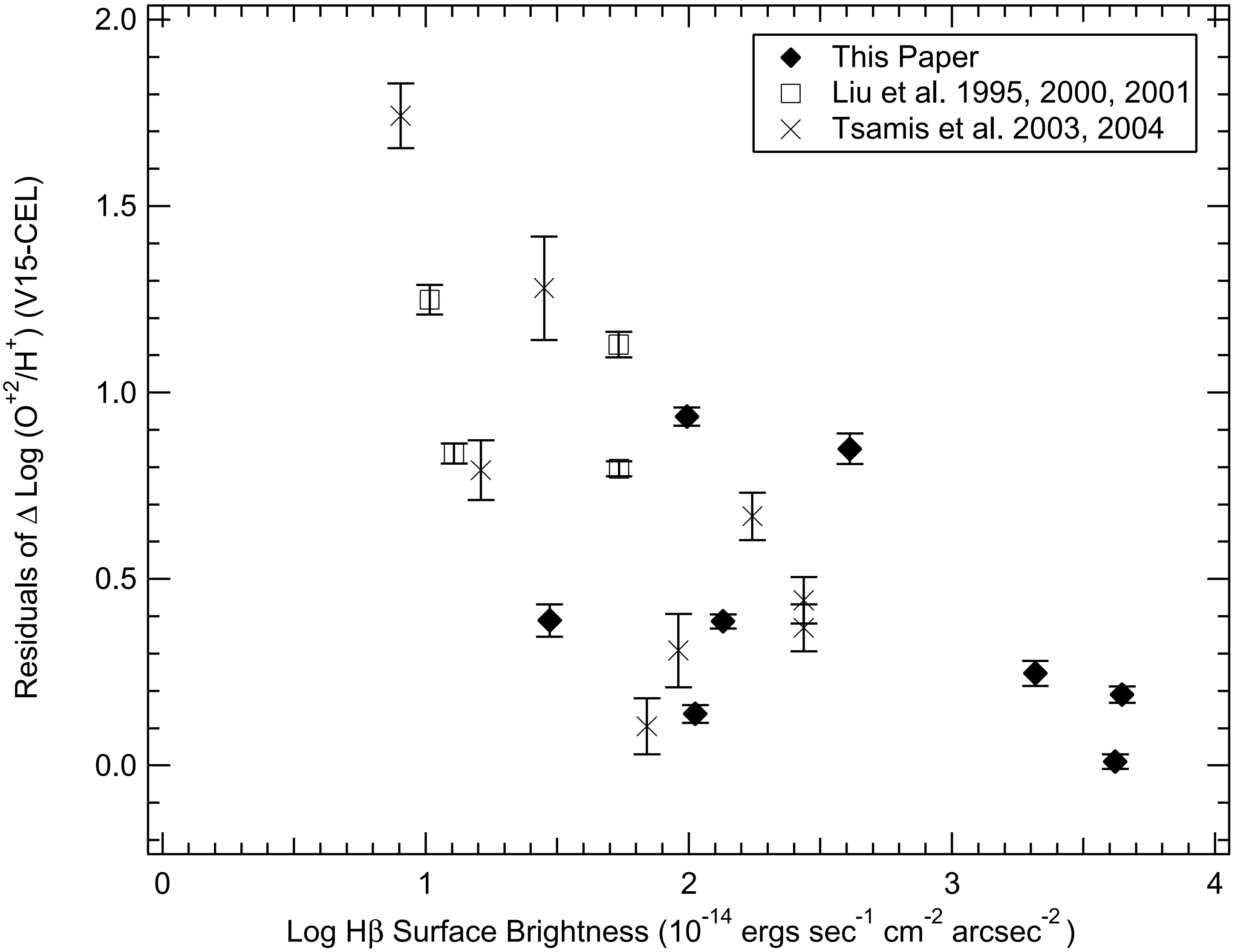

This Paper
Liu et al. 1995, 2000, 2001
Tsamis et al. 2003, 2004
2.0
1.5
1.0
0.5
0.0
0
1
2
3
4
Residuals of Δ Log (O$^{+2}$/H$^{+}$) (V15-CEL)
Log Hβ Surface Brightness (10$^{-14}$ ergs sec$^{-1}$ cm$^{-2}$ arcsec$^{-2}$)

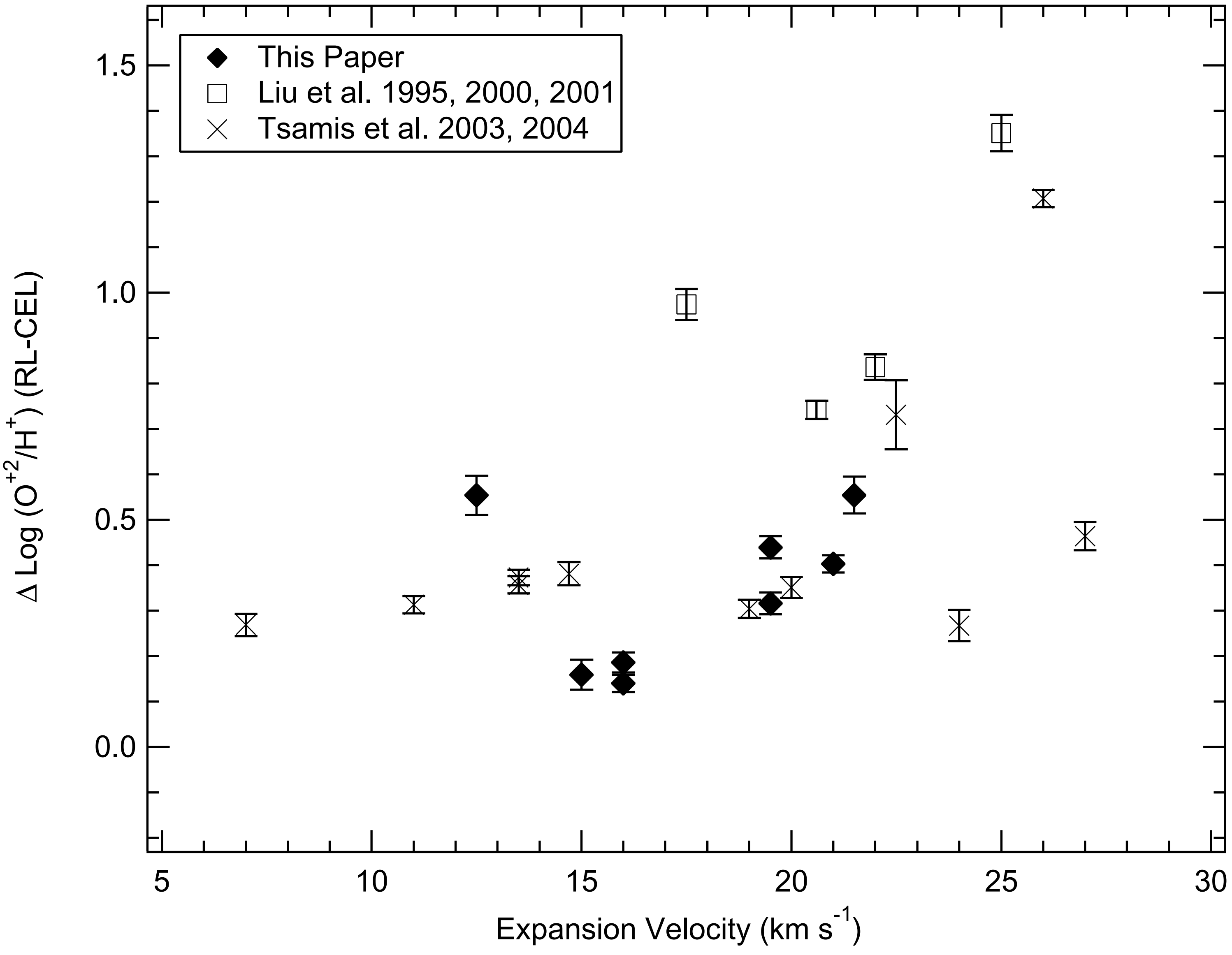
This Paper
Liu et al. 1995, 2000, 2001
Tsamis et al. 2003, 2004
Δ Log (O$^{+2}$/H$^{+}$) (RL-CEL)
Expansion Velocity (km s$^{-1}$)
1.5
1.0
0.5
0.0
5
10
15
20
25
30

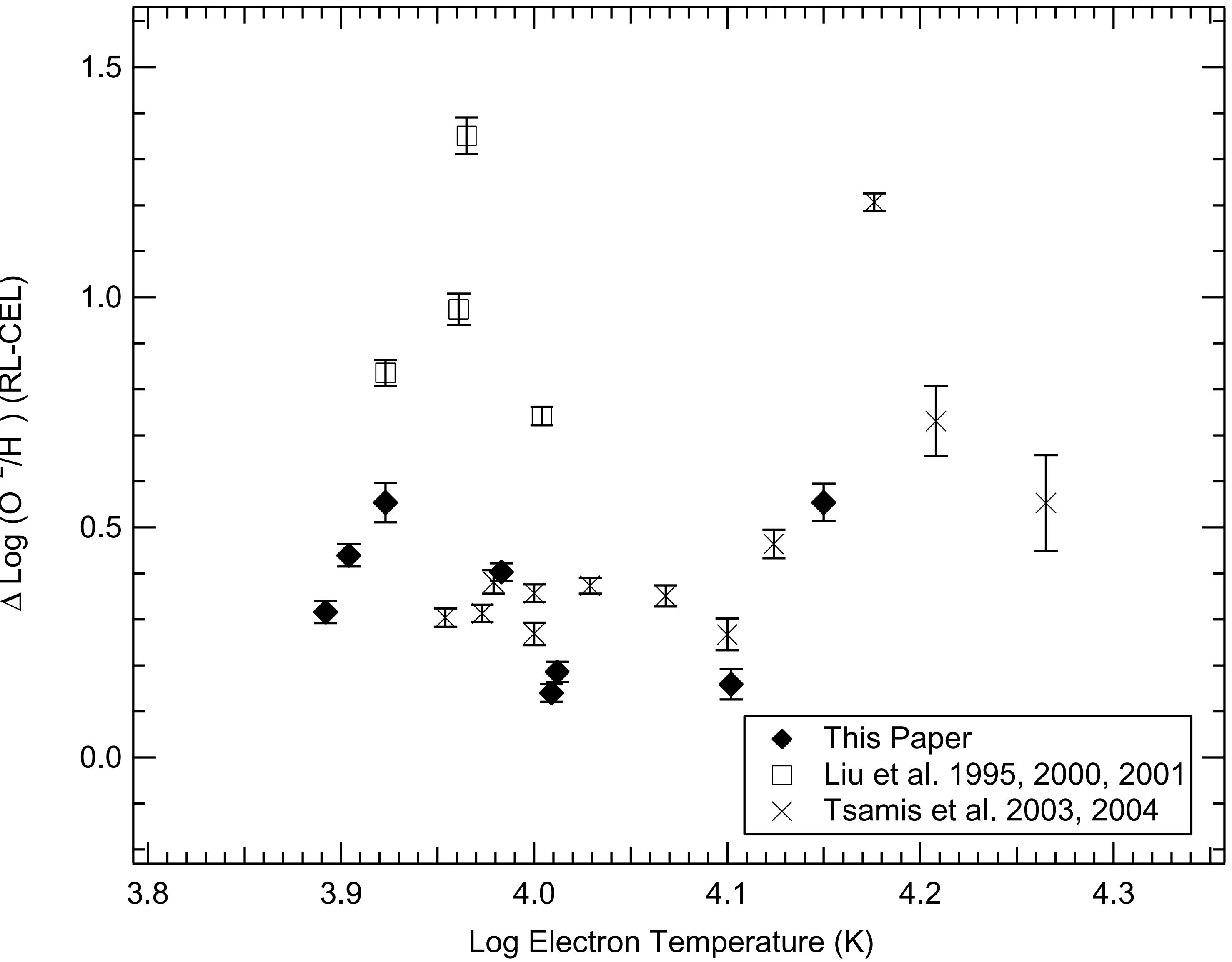

1.5
1.0
0.5
0.0
Δ Log (O+2/H+) (RL-CEL)
3.8
3.9
4.0
4.1
4.2
4.3
Log Electron Temperature (K)
This Paper
Liu et al. 1995, 2000, 2001
Tsamis et al. 2003, 2004

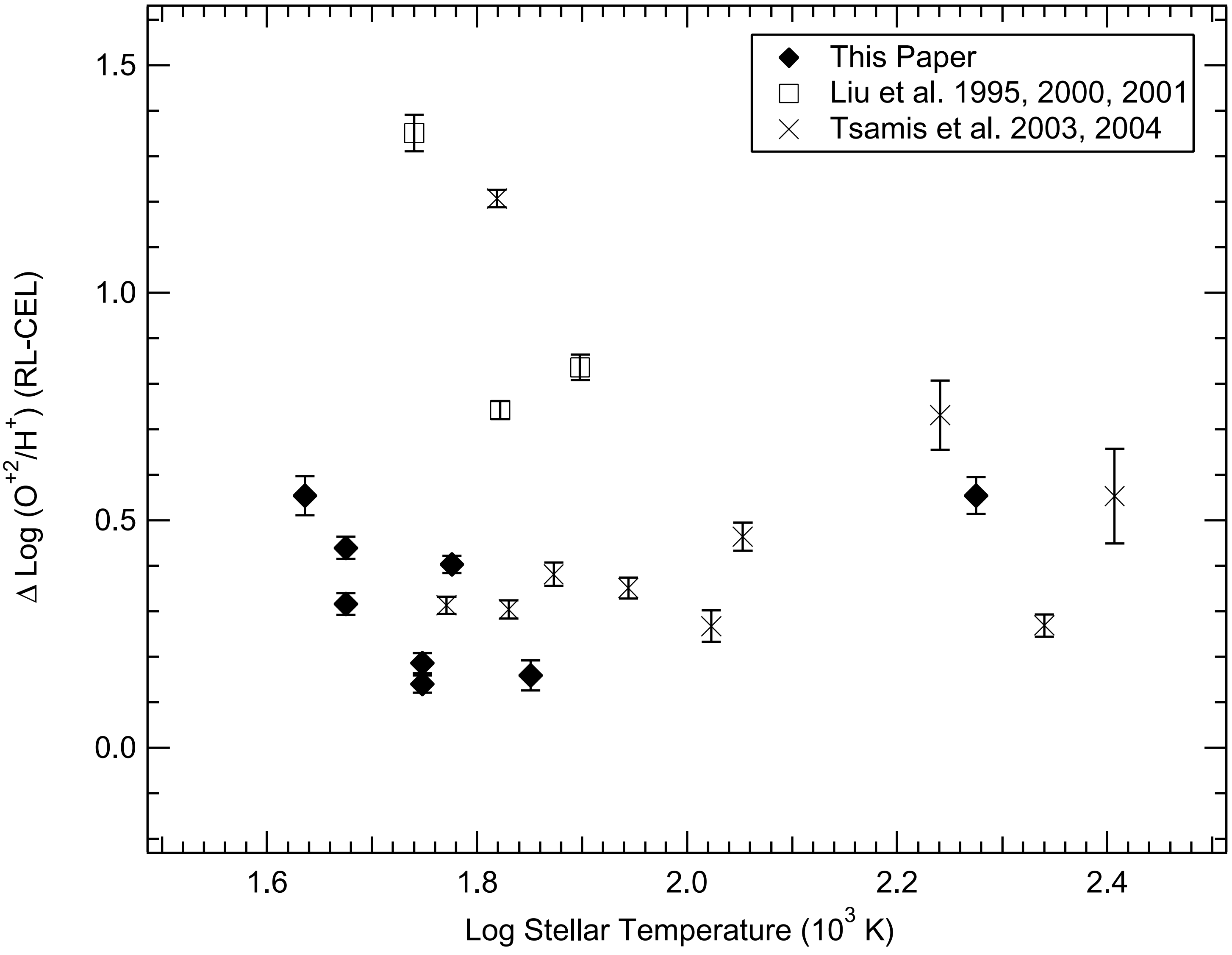

This Paper
Liu et al. 1995, 2000, 2001
Tsamis et al. 2003, 2004
1.5
1.0
0.5
0.0
1.6
1.8
2.0
2.2
2.4
$\Delta$ Log ($O^{+2}/H^{+}$) (RL-CEL)
Log Stellar Temperature ($10^3$ K)

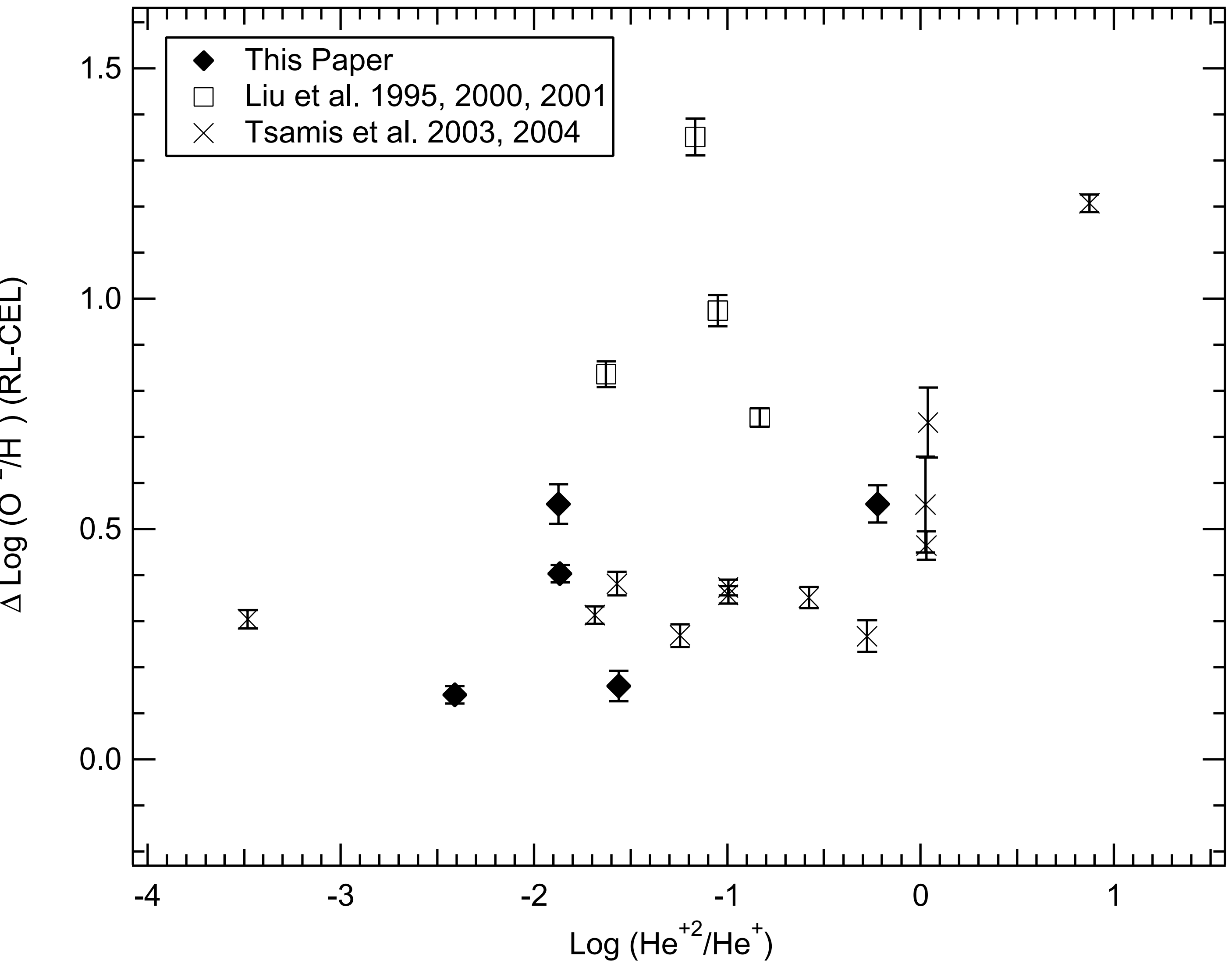

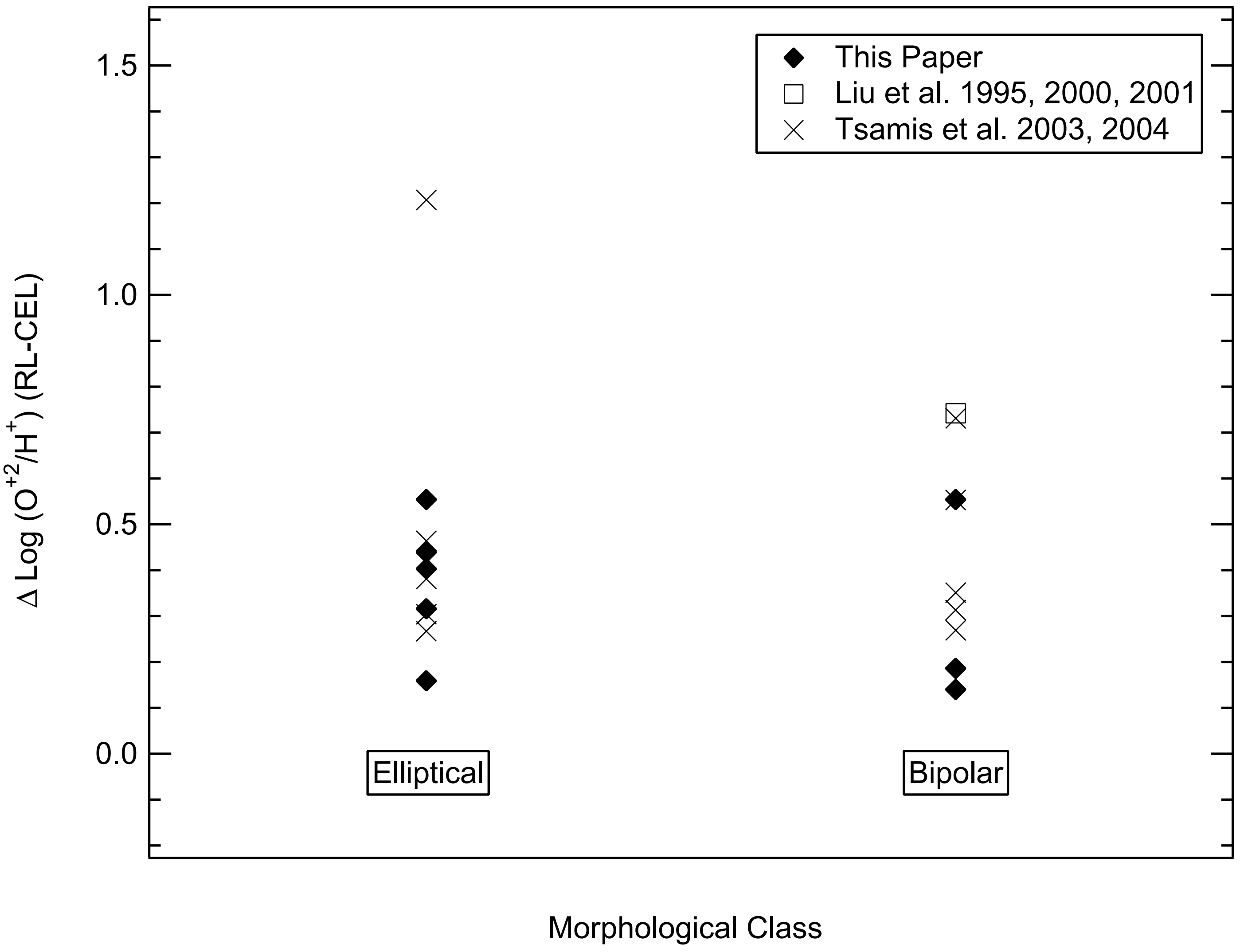

This Paper
Liu et al. 1995, 2000, 2001
Tsamis et al. 2003, 2004
Δ Log (O+2/H+) (RL-CEL)
1.5
1.0
0.5
0.0
Elliptical
Bipolar
Morphological Class

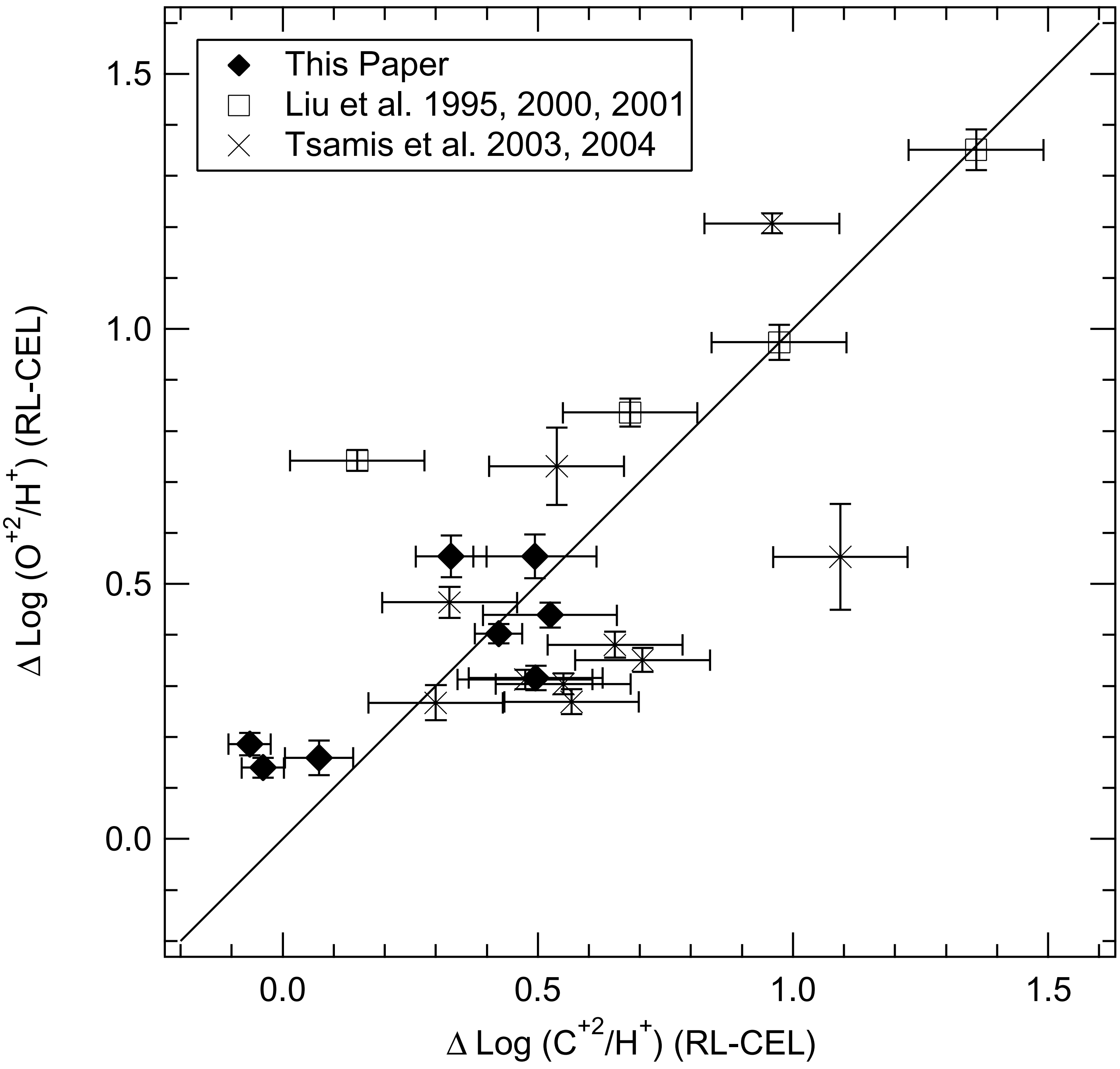
This Paper
Liu et al. 1995, 2000, 2001
Tsamis et al. 2003, 2004
1.5
1.0
0.5
0.0
Δ Log (O+2/H+) (RL-CEL)
0.0
0.5
1.0
1.5
Δ Log (C+2/H+) (RL-CEL)

## TABLE 1

**Objects and Physical Parameters**

| Nebula | Angular Diameter (as) | -Log (F$H\beta$)[i] | Distance (kpc)[j] | Expansion Velocity (km/s)[k] | Log $He^{+2}/He^{+}$ | Stellar Temperature ($x10^3$ K)[l] | c($H\beta$) |
|---|---|---|---|---|---|---|---|
| IC 4593[a] | 13.0[g] | 10.58 | 3.04 | 12.5 | -1.874 | 43 | 0.17 |
| NGC 6210[a] | 16.2[g] | 10.09 | 1.80 | 21.0 | -1.866 | 60 | 0.53 |
| NGC 6543[a] | 19.5[g] | 9.61 | 1.06 | 19.5 | … | 47 | 0.08 |
| NGC 6572[a] | 7.7[h] | 9.82 | 0.91 | 16.0 | -2.410 | 56 | 1.22 |
| NGC 6790[a] | 1.5[h] | 10.90 | 2.36 | 15.0 | -1.561 | 71 | 0.45 |
| NGC 7027[a] | 14.0[g] | 10.12 | 0.65 | 21.5 | -0.223 | 188 | 0.92 |
| M1-42[b] | 12.2[f] | 11.62 | 5.04 | 25.0 | -1.166 | 55 | 0.70 |
| M2-36[b] | 8.0[b] | 11.45 | 6.47 | 22.0 | -1.627 | 79 | 0.27 |
| NGC 6153[c] | 26.0[f] | 10.84 | 1.39 | 17.5 | -1.049 | … | 1.30 |
| NGC 7009[d] | 24.8[f] | 9.78 | 1.10 | 20.6 | -0.833 | 66 | 0.20 |
| NGC 2022[e] | 26.7[f] | 11.13 | 2.49 | 26.0 | 1.169 | 66 | 0.42 |
| NGC 2440[e] | 38.4[f] | 10.50 | 1.54 | 22.5 | 0.298 | 174 | 0.47 |
| NGC 3132[e] | 30.0[h] | 10.45 | 1.13 | 14.7 | -1.530 | 75 | 0.30 |
| NGC 3242[e] | 31.1[f] | 9.79 | 0.95 | 20.0 | -0.550 | 81 | 0.17 |
| NGC 3918[e] | 17.9[f] | 10.04 | 1.22 | 24.0 | -0.314 | 106 | 0.40 |
| NGC 5315[e] | 9.9[f] | 10.42 | 1.92 | 36.0 | -3.153 | 68 | 0.55 |
| NGC 5882[e] | 14.2[f] | 10.38 | 1.69 | 11.0 | -1.657 | 59 | 0.42 |
| NGC 6302[e] | 55.9[f] | 10.55 | 0.55 | … | 0.061 | 255 | 1.39 |
| NGC 6818[e] | 24.7[f] | 10.48 | 1.84 | 27.0 | 0.331 | 113 | 0.37 |
| IC 4191[e] | 4.9[f] | 10.99 | 2.28 | 13.5 | -0.883 | … | 0.70 |
| IC 4406[e] | 35.0[h] | 10.75 | 2.11 | 7.0 | -0.851 | 219 | 0.27 |

[a] This Paper
[b] Liu et al. 2001
[c] Liu et al. 2000
[d] Liu et al. 1995
[e] Tsamis et al. 2003, 2004
[f] Tylenda et al. 2003
[g] Acker et al. 1992
[h] HST images
[i] Cahn et al. 1992
[j] Average of Zhang 1995, Van de Steene & Zijlstra 1995, Bensby & Lundström 2001
[k] Weinberger 1989, Gesicki & Zijlstra 2000 (M2-36)
[l] Average of Preite-Martinez & Pottasch 1983, Preite-Martinez et al. 1989, Kaler & Jacoby 1991

**TABLE 2**

**Observed Fluxes relative to H-Beta=100.0**

**IC 4593**

| λ Obs. | λ Lab | Ident. | Multiplet | Obs. Flux | Corr. Flux | Error |
|---|---|---|---|---|---|---|
| 4187.99 | 4185.45 | O II | V36 | 0.106 | 0.115 | F |
| * | 4186.90 | C III | V18 | * | * | * |
| 4267.97 | 4267.15 | C II | V6 | 0.180 | 0.193 | D |
| 4276.04 | 4273.10 | O II | V67a | 0.134 | 0.144 | C |
| * | 4275.55 | O II | V67a | * | * | * |
| * | 4275.99 | O II | V67b | * | * | * |
| * | 4276.28 | O II | V67b | * | * | * |
| * | 4276.75 | O II | V67b | * | * | * |
| * | 4277.43 | O II | V67c | * | * | * |
| * | 4277.89 | O II | V67b | * | * | * |
| 4286.82 | 4281.32 | O II | V53b | 0.072 | 0.077 | E |
| * | 4282.96 | O II | V67c | * | * | * |
| * | 4283.73 | O II | V67c | * | * | * |
| * | 4285.69 | O II | V78b | * | * | * |
| 4292.15 | 4291.25 | O II | V55 | 0.049 | 0.052 | E |
| * | 4292.21 | O II | V78c | * | * | * |
| 4295.26 | 4294.78 | O II | V53b | 0.034 | 0.036 | F |
| * | 4294.92 | O II | V53b | * | * | * |
| 4304.73 | 4303.82 | O II | V53a | 0.088 | 0.094 | D |
| * | 4303.61 | O II | V65a | * | * | * |
| 4318.43 | 4317.14 | O II | V2 | 0.134 | 0.143 | C |
| * | 4317.70 | O II | V53a | * | * | * |
| * | 4319.63 | O II | V2 | * | * | * |
| 4326.87 | 4325.76 | O II | V2 | 0.163 | 0.173 | C |
| 4341.60 | 4340.47 | H 5 | H 5 | 43.92 | 46.64 | A |
| 4350.53 | 4349.43 | O II | V2 | 0.081 | 0.086 | D |
| 4364.38 | 4363.21 | [O III] | F2 | 1.722 | 1.822 | A |
| 4380.30 | 4379.11 | N III | V18 | 0.125 | 0.132 | C |
| 4389.10 | 4387.93 | He I | V51 | 0.571 | 0.603 | B |
| 4409.80 | 4409.30 | Ne II | V55e | 0.157 | 0.165 | C |
| 4416.98 | 4413.22 | O II | V65 | 0.315 | 0.332 | B |
| * | 4413.11 | O II | V57c | * | * | * |
| * | 4413.11 | O II | V65 | * | * | * |
| * | 4414.90 | O II | V5 | * | * | * |
| * | 4416.97 | O II | V5 | * | * | * |
| 4449.38 | 4448.19 | O II | V35 | 0.067 | 0.070 | F |
| 4454.45 | 4452.36 | O II | V5 | 0.158 | 0.165 | D |
| 4459.95 | 4457.05 | Ne II | V61d | 0.088 | 0.092 | E |
| * | 4457.24 | Ne II | V61d | * | * | * |
| 4472.61 | 4471.49 | He I | V14 | 4.598 | 4.805 | A |
| 4492.87 | 4491.23 | O II | V86a | 0.052 | 0.054 | E |
| * | 4491.07 | C II | | * | * | * |
| 4592.10 | 4590.97 | O II | V15 | 0.046 | 0.048 | E |
| 4597.70 | 4595.96 | O II | V15 | 0.038 | 0.039 | E |
| * | 4596.18 | O II | V15 | * | * | * |
| 4622.48 | 4621.39 | N II | V5 | 0.034 | 0.035 | F |
| 4635.45 | 4634.14 | N III | V2 | 0.577 | 0.592 | B |
| 4642.16 | 4640.64 | N III | V2 | 1.351 | 1.385 | A |
| * | 4641.81 | O II | V1 | * | * | * |
| * | 4641.84 | N III | V2 | * | * | * |
| * | 4643.08 | N II | V5 | * | * | * |
| 4649.18 | 4647.42 | C III | V1 | 1.072 | 1.099 | A |
| * | 4649.13 | O II | V1 | * | * | * |
| 4651.80 | 4650.25 | C III | V1 | 0.935 | 0.958 | B |
| * | 4650.84 | O II | V1 | * | * | * |
| 4659.35 | 4658.26 | [Fe III] | F3 | 0.467 | 0.478 | B |
| 4662.52 | 4661.68 | O II | V1 | 0.107 | 0.110 | E |
| 4667.07 | 4669.27 | O II | V89b | 0.053 | 0.055 | F |
| 4677.12 | 4676.24 | O II | V1 | 0.124 | 0.126 | D |
| 4687.12 | 4685.68 | He II | 3.4 | 1.392 | 1.420 | A |
| 4702.37 | 4701.62 | [Fe III] | F3 | 0.089 | 0.091 | E |
| 4712.15 | 4711.37 | [Ar IV] | F1 | 0.536 | 0.546 | A |
| * | 4713.17 | He I | V12 | * | * | * |
| 4741.70 | 4740.17 | [Ar IV] | F1 | 0.075 | 0.076 | D |
| 4862.45 | 4861.33 | H 4 | H 4 | 100.0 | 100.0 | A |
| 4882.24 | 4881.11 | [Fe III] | F2 | 0.129 | 0.129 | D |
| 4891.65 | 4890.86 | O II | V28 | 0.039 | 0.039 | F |
| 4907.74 | 4906.83 | O II | V28 | 0.159 | 0.158 | D |
| 4923.06 | 4921.93 | He I | V48 | 1.433 | 1.427 | A |
| 4932.53 | 4931.80 | [O III] | F1 | 0.088 | 0.087 | D |
| 4960.03 | 4958.91 | [O III] | F1 | 185.6 | 184.1 | A |

**NGC 6210**

| λ Obs. | λ Lab | Ident. | Multiplet | Obs. Flux | Corr. Flux | Error |
|---|---|---|---|---|---|---|
| 4168.84 | 4168.97 | He I | V52 | 0.055 | 0.070 | B |
| * | 4169.22 | O II | V19 | * | * | * |
| 4185.91 | 4185.45 | O II | V36 | 0.063 | 0.080 | B |
| * | 4186.90 | C III | V18 | * | * | * |
| 4195.16 | 4195.76 | N III | V6 | 0.020 | 0.025 | C |
| 4199.62 | 4199.83 | He II | 4.11 | 0.053 | 0.067 | B |
| * | 4200.10 | N III | V6 | * | * | * |
| 4219.11 | 4219.37 | Ne II | V52a | 0.041 | 0.051 | B |
| * | 4219.74 | Ne II | V52a | * | * | * |
| 4227.42 | 4227.74 | N II | V33 | 0.007 | 0.009 | F |
| * | 4227.20 | [Fe V] | F2 | * | * | * |
| 4237.18 | 4236.91 | N II | V48a | 0.013 | 0.016 | D |
| * | 4237.05 | N II | V48b | * | * | * |
| 4241.52 | 4241.24 | N II | V48a | 0.015 | 0.019 | D |
| * | 4241.78 | N II | V48b | * | * | * |
| 4253.64 | 4254.00 | O II | V101 | 0.029 | 0.036 | C |
| 4266.81 | 4267.15 | C II | V6 | 0.284 | 0.350 | A |
| 4275.87 | 4273.10 | O II | V67a | 0.116 | 0.142 | A |
| * | 4275.55 | O II | V67a | * | * | * |
| * | 4275.99 | O II | V67b | * | * | * |
| * | 4276.28 | O II | V67b | * | * | * |
| * | 4276.75 | O II | V67b | * | * | * |
| * | 4277.43 | O II | V67c | * | * | * |
| * | 4277.89 | O II | V67b | * | * | * |
| 4283.70 | 4281.32 | O II | V53b | 0.039 | 0.048 | C |
| * | 4282.96 | O II | V67c | * | * | * |
| * | 4283.73 | O II | V67c | * | * | * |
| * | 4285.69 | O II | V78b | * | * | * |
| 4291.10 | 4291.25 | O II | V55 | 0.028 | 0.035 | C |
| * | 4292.21 | O II | V78c | * | * | * |
| 4294.60 | 4294.78 | O II | V53b | 0.033 | 0.041 | C |
| * | 4294.92 | O II | V53b | * | * | * |
| 4303.44 | 4303.82 | O II | V53a | 0.067 | 0.082 | B |
| * | 4303.61 | O II | V65a | * | * | * |
| 4317.53 | 4317.14 | O II | V2 | 0.102 | 0.124 | B |
| * | 4317.70 | O II | V53a | * | * | * |
| * | 4319.63 | O II | V2 | * | * | * |
| 4325.18 | 4325.76 | O II | V2 | 0.015 | 0.018 | E |
| 4340.09 | 4340.47 | H 5 | H 5 | 38.90 | 46.80 | A |
| 4349.08 | 4349.43 | O II | V2 | 0.062 | 0.074 | C |
| 4362.84 | 4363.21 | [O III] | F2 | 5.187 | 6.172 | A |
| 4378.70 | 4379.11 | N III | V18 | 0.092 | 0.108 | B |
| 4387.59 | 4387.93 | He I | V51 | 0.534 | 0.629 | A |
| 4391.46 | 4391.99 | Ne II | V55e | 0.039 | 0.046 | C |
| * | 4392.00 | Ne II | V55e | * | * | * |
| 4408.80 | 4409.30 | Ne II | V55e | 0.020 | 0.024 | D |
| 4415.42 | 4413.22 | Ne II | V65 | 0.101 | 0.118 | B |
| * | 4413.11 | Ne II | V57c | * | * | * |
| * | 4413.11 | Ne II | V65 | * | * | * |
| * | 4414.90 | O II | V5 | * | * | * |
| * | 4416.97 | O II | V5 | * | * | * |
| 4428.31 | 4428.64 | Ne II | V60c | 0.020 | 0.023 | C |
| * | 4428.52 | Ne II | V61b | * | * | * |
| 4430.97 | 4431.82 | N II | V55a | 0.018 | 0.021 | C |
| * | 4432.74 | N II | V55a | * | * | * |
| * | 4433.48 | N II | V55b | * | * | * |
| 4437.19 | 4437.55 | He I | V50 | 0.056 | 0.064 | B |
| 4447.92 | 4448.19 | O II | V35 | 0.012 | 0.014 | D |
| 4452.47 | 4452.36 | O II | V5 | 0.010 | 0.011 | D |
| 4456.82 | 4457.05 | Ne II | V61d | 0.011 | 0.013 | D |
| * | 4457.24 | Ne II | V61d | * | * | * |
| 4471.15 | 4471.49 | He I | V14 | 4.539 | 5.198 | A |
| 4479.99 | 4481.21 | Mg II | V4 | 0.025 | 0.029 | C |
| 4487.82 | 4487.72 | O II | V104 | 0.017 | 0.019 | D |
| * | 4488.20 | O II | V104 | * | * | * |
| * | 4489.49 | O II | V86b | * | * | * |
| 4490.89 | 4491.23 | O II | V86a | 0.023 | 0.026 | C |
| * | 4491.07 | C II | | * | * | * |
| 4498.18 | 4498.92 | Ne II | V64c | 0.010 | 0.011 | E |
| * | 4499.12 | Ne II | V64c | * | * | * |
| 4510.43 | 4510.91 | N III | V3 | 0.043 | 0.049 | B |
| 4514.56 | 4514.86 | N III | V3 | 0.013 | 0.015 | D |
| 4517.68 | 4518.15 | N III | V3 | 0.017 | 0.019 | D |
| 4523.06 | 4523.58 | N III | V3 | 0.016 | 0.018 | D |
| 4529.96 | 4530.41 | N II | V58b | 0.014 | 0.016 | D |
| * | 4530.86 | N III | V3 | * | * | * |
| 4534.55 | 4534.58 | N III | V3 | 0.010 | 0.011 | E |
| 4540.98 | 4539.71 | N III | V12 | 0.031 | 0.034 | C |
| * | 4541.59 | He II | 4.9 | * | * | * |
| 4544.67 | 4544.85 | N III | V12 | 0.007 | 0.008 | F |
| 4552.56 | 4552.53 | N II | V58a | 0.017 | 0.019 | D |
| 4562.39 | 4562.60 | Mg I] | | 0.024 | 0.027 | C |
| 4570.96 | 4571.10 | Mg I] | | 0.102 | 0.113 | B |
| 4590.62 | 4590.97 | O II | V15 | 0.057 | 0.062 | B |
| 4595.82 | 4595.96 | O II | V15 | 0.039 | 0.043 | C |
| * | 4596.18 | O II | V15 | * | * | * |
| 4601.04 | 4601.48 | N II | V5 | 0.019 | 0.021 | D |
| * | 4602.13 | O II | V92b | * | * | * |
| 4609.05 | 4607.16 | N II | V5 | 0.093 | 0.101 | B |
| * | 4609.44 | O II | V92a | * | * | * |
| 4614.20 | 4613.14 | O II | V92b | 0.023 | 0.025 | D |
| * | 4613.68 | O II | V92b | * | * | * |
| 4621.17 | 4621.39 | N II | V5 | 0.011 | 0.012 | E |
| * | 4630.54 | N II | V5 | * | * | * |
| 4633.70 | 4634.14 | N III | V2 | 0.279 | 0.303 | A |
| 4640.45 | 4640.64 | N III | V2 | 0.965 | 1.043 | A |
| * | 4641.81 | O II | V1 | * | * | * |
| * | 4641.84 | N III | V2 | * | * | * |
| * | 4643.08 | N II | V5 | * | * | * |
| 4649.83 | 4647.42 | C III | V1 | 0.628 | 0.677 | A |
| * | 4649.13 | O II | V1 | * | * | * |
| * | 4650.25 | C III | V1 | * | * | * |
| * | 4650.84 | O II | V1 | * | * | * |
| 4657.94 | 4658.26 | [Fe III] | F3 | 0.117 | 0.125 | A |
| 4661.21 | 4661.68 | O II | V1 | 0.136 | 0.146 | A |
| 4668.44 | 4669.22 | O II | V89b | 0.003 | 0.003 | F |
| 4675.52 | 4676.24 | O II | V1 | 0.122 | 0.130 | A |
| 4685.18 | 4685.68 | He II | 3.4 | 1.471 | 1.567 | A |
| 4695.91 | 4696.35 | O II | V1 | 0.013 | 0.014 | D |
| 4698.79 | 4699.22 | O II | V25 | 0.015 | 0.016 | D |
| 4701.40 | 4701.62 | [Fe III] | F3 | 0.037 | 0.039 | C |
| 4705.25 | 4705.35 | O II | V25 | 0.014 | 0.014 | D |
| 4711.98 | 4711.37 | [Ar IV] | F1 | 1.909 | 2.015 | A |
| * | 4713.17 | He I | V12 | * | * | * |
| 4739.76 | 4740.17 | [Ar IV] | F1 | 1.431 | 1.497 | A |
| 4773.14 | 4772.93 | Ne II | | 0.005 | 0.006 | E |
| 4802.29 | 4802.23 | C II | | 0.013 | 0.013 | C |
| * | 4803.29 | N II | V20 | * | * | * |
| 4814.46 | 4815.55 | S II | V9 | 0.010 | 0.010 | D |
| 4860.89 | 4861.33 | H 4 | H 4 | 100.0 | 100.0 | A |
| 4880.65 | 4881.11 | [Fe III] | F2 | 0.057 | 0.057 | C |
| 4890.18 | 4890.86 | O II | V28 | 0.013 | 0.013 | F |
| 4906.18 | 4906.83 | O II | V28 | 0.045 | 0.044 | D |
| 4921.52 | 4921.93 | He I | V48 | 1.317 | 1.300 | A |
| 4930.77 | 4931.80 | [O III] | F1 | 0.125 | 0.123 | C |
| 4958.48 | 4958.91 | [O III] | F1 | 381.5 | 372.4 | A |

**NGC 6543**

| λ Obs. | λ Lab | Ident. | Multiplet | Obs. Flux | Corr. Flux | Error |
|---|---|---|---|---|---|---|
| 4167.51 | 4168.97 | He I | V52 | 0.093 | 0.096 | D |
| * | 4169.22 | O II | V19 | * | * | * |
| 4183.99 | 4185.45 | O II | V36 | 0.057 | 0.059 | E |
| * | 4186.90 | C III | V18 | * | * | * |
| 4193.37 | 4195.76 | N III | V6 | 0.029 | 0.030 | F |
| 4198.19 | 4199.83 | He II | 4.11 | 0.121 | 0.126 | D |
| * | 4200.10 | N III | V6 | * | * | * |
| 4219.17 | 4219.37 | Ne II | V52a | 0.086 | 0.089 | D |
| * | 4219.74 | Ne II | V52a | * | * | * |
| 4226.86 | 4227.74 | N II | V33 | 0.050 | 0.051 | E |
| * | 4227.20 | [Fe V] | F2 | * | * | * |
| 4230.47 | 4231.53 | Ne II | V52b | 0.057 | 0.059 | E |
| * | 4231.64 | Ne II | V52b | * | * | * |
| 4235.67 | 4236.91 | N II | V48a | 0.058 | 0.060 | E |
| * | 4237.05 | N II | V48b | * | * | * |
| 4240.00 | 4241.24 | N II | V48a | 0.071 | 0.074 | E |
| * | 4241.78 | N II | V48b | * | * | * |
| 4252.35 | 4254.00 | O II | V101 | 0.033 | 0.034 | E |
| 4265.67 | 4267.15 | C II | V6 | 0.635 | 0.655 | A |
| 4274.55 | 4273.10 | O II | V67a | 0.124 | 0.127 | B |
| * | 4275.55 | O II | V67a | * | * | * |
| * | 4275.99 | O II | V67b | * | * | * |
| * | 4276.28 | O II | V67b | * | * | * |
| * | 4276.75 | O II | V67b | * | * | * |
| * | 4277.43 | O II | V67c | * | * | * |
| * | 4277.89 | O II | V67b | * | * | * |
| 4282.81 | 4281.32 | O II | V53b | 0.049 | 0.050 | D |
| * | 4282.96 | O II | V67c | * | * | * |
| * | 4283.73 | O II | V67c | * | * | * |
| * | 4285.69 | O II | V78b | * | * | * |
| 4289.80 | 4291.25 | O II | V55 | 0.047 | 0.048 | E |
| * | 4292.21 | O II | V78c | * | * | * |
| 4293.61 | 4294.78 | O II | V53b | 0.052 | 0.053 | E |
| * | 4294.92 | O II | V53b | * | * | * |
| 4302.43 | 4303.82 | O II | V53a | 0.088 | 0.091 | C |
| * | 4303.61 | O II | V65a | * | * | * |
| 4315.98 | 4317.14 | O II | V2 | 0.147 | 0.151 | C |
| * | 4317.70 | O II | V53a | * | * | * |
| * | 4319.63 | O II | V2 | * | * | * |
| 4323.87 | 4325.76 | O II | V2 | 0.021 | 0.021 | F |
| 4339.00 | 4340.47 | H 5 | H 5 | 45.34 | 46.59 | A |
| 4347.20 | 4349.43 | O II | V2 | 0.360 | 0.370 | A |
| 4361.73 | 4363.21 | [O III] | F2 | 1.902 | 1.951 | A |
| 4377.44 | 4379.11 | N III | V18 | 0.107 | 0.110 | C |
| 4386.39 | 4387.93 | He I | V51 | 0.725 | 0.743 | A |
| 4390.39 | 4391.99 | Ne II | V55e | 0.054 | 0.055 | C |
| * | 4392.00 | Ne II | V55e | * | * | * |
| 4407.68 | 4409.30 | Ne II | V55e | 0.043 | 0.044 | D |
| 4414.36 | 4413.22 | Ne II | V65 | 0.139 | 0.142 | B |
| * | 4413.11 | Ne II | V57c | * | * | * |
| * | 4413.11 | Ne II | V65 | * | * | * |
| * | 4414.90 | O II | V5 | * | * | * |
| * | 4416.97 | O II | V5 | * | * | * |
| 4426.84 | 4428.64 | Ne II | V60c | 0.044 | 0.045 | E |
| * | 4428.52 | Ne II | V61b | * | * | * |
| 4430.54 | 4431.82 | N II | V55a | 0.035 | 0.036 | F |
| * | 4432.74 | N II | V55a | * | * | * |
| * | 4433.48 | N II | V55b | * | * | * |
| 4436.02 | 4437.55 | He I | V50 | 0.059 | 0.060 | E |
| 4451.65 | 4452.36 | O II | V5 | 0.037 | 0.038 | F |
| 4456.34 | 4457.05 | Ne II | V61d | 0.029 | 0.030 | F |
| * | 4457.24 | Ne II | V61d | | | * |
| 4469.97 | 4471.49 | He I | V14 | 6.005 | 6.126 | A |
| 4478.90 | 4481.21 | Mg II ? | V4 | 0.039 | 0.040 | F |
| 4486.44 | 4487.72 | O II | V104 | 0.028 | 0.029 | E |
| * | 4488.20 | O II | V104 | * | * | * |
| * | 4489.49 | O II | V86b | * | * | * |
| 4489.33 | 4491.23 | O II | V86a | 0.037 | 0.038 | D |
| * | 4491.07 | C II | | * | * | * |
| 4509.09 | 4510.91 | N III | V3 | 0.075 | 0.076 | C |
| 4514.55 | 4514.86 | N III | V3 | 0.034 | 0.034 | E |
| 4517.41 | 4518.15 | N III | V3 | 0.075 | 0.076 | C |
| 4522.34 | 4523.58 | N III | V3 | 0.023 | 0.023 | E |
| 4528.60 | 4530.41 | N II | V58b | 0.049 | 0.049 | D |
| * | 4530.86 | N III | V3 | * | * | * |
| 4543.34 | 4544.85 | N III | V12 | 0.017 | 0.018 | F |
| 4552.07 | 4552.53 | N II | V58a | 0.080 | 0.081 | C |
| 4560.90 | 4562.60 | Mg I] | | 0.016 | 0.017 | E |
| 4569.63 | 4571.10 | Mg I] | | 0.041 | 0.041 | C |
| 4589.54 | 4590.97 | O II | V15 | 0.118 | 0.120 | B |
| 4594.55 | 4595.96 | O II | V15 | 0.115 | 0.116 | B |
| * | 4596.18 | O II | V15 | * | * | * |
| 4605.78 | 4607.16 | N II | V5 | 0.768 | 0.779 | A |
| * | 4609.44 | O II | V92a | * | * | * |
| 4619.51 | 4621.39 | N II | V5 | 0.232 | 0.235 | A |
| 4629.15 | 4630.54 | N II | V5 | 0.230 | 0.232 | B |
| 4632.50 | 4634.14 | N III | V2 | 0.718 | 0.727 | A |
| 4639.35 | 4640.64 | N III | V2 | 1.795 | 1.816 | A |
| * | 4641.81 | O II | V1 | * | * | * |
| * | 4641.84 | N III | V2 | * | * | * |
| * | 4643.08 | N II | V5 | * | * | * |
| 4648.02 | 4647.42 | C III | V1 | 0.885 | 0.895 | A |
| * | 4649.13 | O II | V1 | * | * | * |
| * | 4650.25 | C III | V1 | * | * | * |
| * | 4650.84 | O II | V1 | * | * | * |
| 4656.24 | 4658.26 | [Fe III] | F3 | 0.431 | 0.436 | B |
| 4659.87 | 4661.68 | O II | V1 | 0.238 | 0.241 | B |
| 4674.68 | 4676.24 | O II | V1 | 0.117 | 0.119 | C |
| 4685.22 | 4685.68 | He II | 3.4 | 3.627 | 3.661 | A |
| 4705.30 | 4705.35 | O II | V25 | 0.019 | 0.019 | F |
| 4712.05 | 4711.37 | [Ar IV] | F1 | 1.720 | 1.734 | A |
| * | 4713.17 | He I | V12 | * | * | * |
| 4738.64 | 4740.17 | [Ar IV] | F1 | 1.105 | 1.113 | A |
| 4771.48 | 4772.93 | Ne II | | 0.020 | 0.020 | E |
| 4777.82 | 4779.72 | N II | V20 | 0.016 | 0.016 | F |
| 4801.33 | 4802.23 | C II | | 0.077 | 0.077 | C |
| * | 4803.29 | N II | V20 | * | * | * |
| 4812.90 | 4815.55 | S II | V9 | 0.037 | 0.037 | D |
| 4859.73 | 4861.33 | H 4 | H 4 | 100.0 | 100.0 | A |
| 4879.41 | 4881.11 | [Fe III] | F2 | 0.023 | 0.023 | E |
| 4889.41 | 4890.86 | O II | V28 | 0.031 | 0.031 | D |
| 4905.14 | 4906.83 | O II | V28 | 0.074 | 0.074 | C |
| 4920.30 | 4921.93 | He I | V48 | 1.630 | 1.627 | A |
| 4929.22 | 4931.80 | [O III] | F1 | 0.135 | 0.134 | B |
| 4942.45 | 4943.00 | O II | V33 | 0.078 | 0.078 | C |
| 4957.28 | 4958.91 | [O III] | F1 | 233.6 | 232.8 | A |

**NGC 6543-S**

| λ Obs. | λ Lab | Ident. | Multiplet | Obs. Flux | Corr. Flux | Error |
|---|---|---|---|---|---|---|
| 4167.84 | 4168.97 | He I | V52 | 0.067 | 0.070 | C |
| * | 4169.22 | O II | V19 | * | * | * |
| 4184.35 | 4185.45 | O II | V36 | 0.060 | 0.063 | D |
| * | 4186.90 | C III | V18 | * | * | * |
| 4193.62 | 4195.76 | N III | V6 | 0.024 | 0.025 | F |
| 4218.47 | 4219.37 | Ne II | V52a | 0.042 | 0.044 | E |
| * | 4219.74 | Ne II | V52a | * | * | * |
| 4230.94 | 4231.53 | Ne II | V52b | 0.014 | 0.015 | F |
| * | 4231.64 | Ne II | V52b | * | * | * |
| 4235.66 | 4236.91 | N II | V48a | 0.016 | 0.017 | F |
| * | 4237.05 | N II | V48b | * | * | * |
| 4240.26 | 4241.24 | N II | V48a | 0.035 | 0.037 | F |
| * | 4241.78 | N II | V48b | * | * | * |
| 4252.20 | 4254.00 | O II | V101 | 0.029 | 0.030 | E |
| 4265.69 | 4267.15 | C II | V6 | 0.589 | 0.615 | A |
| 4274.82 | 4273.10 | O II | V67a | 0.105 | 0.109 | C |
| * | 4275.55 | O II | V67a | * | * | * |
| * | 4275.99 | O II | V67b | * | * | * |
| * | 4276.28 | O II | V67b | * | * | * |
| * | 4276.75 | O II | V67b | * | * | * |
| * | 4277.43 | O II | V67c | * | * | * |
| * | 4277.89 | O II | V67b | * | * | * |
| 4282.81 | 4281.32 | O II | V53b | 0.048 | 0.050 | D |
| * | 4282.96 | O II | V67c | * | * | * |
| * | 4283.73 | O II | V67c | * | * | * |
| * | 4285.69 | O II | V78b | * | * | * |
| 4290.04 | 4291.25 | O II | V55 | 0.029 | 0.030 | E |
| * | 4292.21 | O II | V78c | * | * | * |
| 4293.33 | 4294.78 | O II | V53b | 0.027 | 0.028 | E |
| * | 4294.92 | O II | V53b | * | * | * |
| 4302.53 | 4303.82 | O II | V53a | 0.060 | 0.062 | C |
| * | 4303.61 | O II | V65a | * | * | * |
| 4316.44 | 4317.14 | O II | V2 | 0.138 | 0.144 | B |
| * | 4317.70 | O II | V53a | * | * | * |
| * | 4319.63 | O II | V2 | * | * | * |
| 4324.91 | 4325.76 | O II | V2 | 0.023 | 0.024 | E |
| 4338.96 | 4340.47 | H 5 | H 5 | 44.82 | 46.55 | A |
| 4347.89 | 4349.43 | O II | V2 | 0.077 | 0.079 | C |
| 4361.67 | 4363.21 | [O III] | F2 | 1.592 | 1.650 | A |
| 4377.82 | 4379.11 | N III | V18 | 0.074 | 0.077 | C |
| 4386.41 | 4387.93 | He I | V51 | 0.689 | 0.713 | A |
| 4390.54 | 4391.99 | Ne II | V55e | 0.017 | 0.018 | F |
| * | 4392.00 | Ne II | V55e | * | * | * |
| 4407.75 | 4409.30 | Ne II | V55e | 0.044 | 0.046 | D |
| 4414.36 | 4413.22 | Ne II | V65 | 0.129 | 0.134 | C |
| * | 4413.11 | Ne II | V57c | * | * | * |
| * | 4413.11 | Ne II | V65 | * | * | * |
| * | 4414.90 | O II | V5 | * | * | * |
| * | 4416.97 | O II | V5 | * | * | * |
| 4426.75 | 4428.64 | Ne II | V60c | 0.039 | 0.040 | E |
| * | 4428.52 | Ne II | V61b | * | * | * |
| 4430.54 | 4431.82 | N II | V55a | 0.072 | 0.075 | C |
| * | 4432.74 | N II | V55a | * | * | * |
| * | 4433.48 | N II | V55b | * | * | * |
| 4436.09 | 4437.55 | He I | V50 | 0.055 | 0.057 | D |
| 4446.17 | 4448.19 | O II | V35 | 0.015 | 0.015 | F |
| 4451.10 | 4452.36 | O II | V5 | 0.011 | 0.011 | F |
| 4469.99 | 4471.49 | He I | V14 | 5.886 | 6.054 | A |
| 4486.94 | 4487.72 | O II | V104 | 0.024 | 0.024 | D |
| 4489.73 | 4488.20 | O II | V104 | 0.028 | 0.028 | D |
| * | 4489.49 | O II | V86b | * | * | * |
| * | 4491.23 | O II | V86a | * | * | * |
| * | 4491.07 | C II | | * | * | * |
| 4497.21 | 4498.92 | Ne II | V64c | 0.015 | 0.015 | E |
| * | 4499.12 | Ne II | V64c | * | * | * |
| 4528.72 | 4530.41 | N II | V58b | 0.032 | 0.032 | D |
| * | 4530.86 | N III | V3 | * | * | * |
| 4551.22 | 4552.53 | N II | V58a | 0.025 | 0.026 | D |
| 4560.74 | 4562.60 | Mg I] | | 0.017 | 0.017 | E |
| 4569.38 | 4571.10 | Mg I] | | 0.055 | 0.056 | C |
| 4589.54 | 4590.97 | O II | V15 | 0.068 | 0.070 | C |
| 4594.71 | 4595.96 | O II | V15 | 0.049 | 0.050 | C |
| * | 4596.18 | O II | V15 | * | * | * |
| 4600.20 | 4601.48 | N II | V5 | 0.040 | 0.041 | C |
| * | 4602.13 | O II | V92b | * | * | * |
| 4605.78 | 4607.16 | N II | V5 | 0.106 | 0.108 | B |
| * | 4609.44 | O II | V92a | * | * | * |
| 4619.62 | 4621.39 | N II | V5 | 0.037 | 0.038 | D |
| 4629.07 | 4630.54 | N II | V5 | 0.103 | 0.105 | B |
| 4632.73 | 4634.14 | N III | V2 | 0.244 | 0.248 | B |
| 4639.35 | 4640.64 | N III | V2 | 0.973 | 0.989 | A |
| * | 4641.81 | O II | V1 | * | * | * |
| * | 4641.84 | N III | V2 | * | * | * |
| * | 4643.08 | N II | V5 | * | * | * |
| 4648.03 | 4647.42 | C III | V1 | 0.619 | 0.629 | A |
| * | 4649.13 | O II | V1 | * | * | * |
| * | 4650.25 | C III | V1 | * | * | * |
| * | 4650.84 | O II | V1 | * | * | * |
| 4656.74 | 4658.26 | [Fe III] | F3 | 0.334 | 0.339 | B |
| 4660.12 | 4661.68 | O II | V1 | 0.180 | 0.183 | B |
| 4668.44 | 4669.22 | O II | V89b | 0.020 | 0.021 | F |
| 4674.42 | 4676.24 | O II | V1 | 0.157 | 0.159 | B |
| 4684.65 | 4685.68 | He II | 3.4 | 0.022 | 0.022 | F |
| 4700.16 | 4701.62 | [Fe III] | F3 | 0.094 | 0.095 | B |
| 4710.45 | 4711.37 | [Ar IV] | F1 | 1.242 | 1.256 | A |
| * | 4713.17 | He I | V12 | * | * | * |
| 4738.72 | 4740.17 | [Ar IV] | F1 | 0.602 | 0.608 | A |
| 4777.14 | 4779.72 | N II | V20 | 0.038 | 0.038 | C |
| 4786.37 | 4788.13 | N II | V20 | 0.020 | 0.020 | D |
| 4801.46 | 4802.23 | C II | | 0.049 | 0.049 | C |
| * | 4803.29 | N II | V20 | * | * | * |
| 4859.71 | 4861.33 | H 4 | H 4 | 100.0 | 100.0 | A |
| 4879.60 | 4881.11 | [Fe III] | F2 | 0.141 | 0.141 | B |
| 4889.21 | 4890.86 | O II | V28 | 0.023 | 0.023 | E |
| 4905.15 | 4906.83 | O II | V28 | 0.058 | 0.058 | C |
| 4920.33 | 4921.93 | He I | V48 | 1.605 | 1.601 | A |
| 4929.56 | 4931.80 | [O III] | F1 | 0.087 | 0.086 | C |
| 4957.26 | 4958.91 | [O III] | F1 | 224.7 | 223.6 | A |

**NGC 6572**

| λ Obs. | λ Lab | Ident. | Multiplet | Obs. Flux | Corr. Flux | Error |
|---|---|---|---|---|---|---|
| 4169.56 | 4168.97 | He I | V52 | 0.036 | 0.062 | A |
| * | 4169.22 | O II | V19 | * | * | * |
| 4186.94 | 4185.45 | O II | V36 | 0.047 | 0.080 | A |
| * | 4186.90 | C III | V18 | * | * | * |
| 4196.89 | 4195.76 | N III | V6 | 0.009 | 0.016 | C |
| 4200.40 | 4199.83 | He II | 4.11 | 0.027 | 0.046 | B |
| * | 4200.10 | N III | V6 | * | * | * |
| 4220.40 | 4219.37 | Ne II | V52a | 0.015 | 0.025 | C |
| * | 4219.74 | Ne II | V52a | * | * | * |
| 4237.81 | 4236.91 | N II | V48a | 0.007 | 0.012 | D |
| * | 4237.05 | N II | V48b | * | * | * |
| 4242.12 | 4241.24 | N II | V48a | 0.010 | 0.016 | C |
| * | 4241.78 | N II | V48b | * | * | * |
| 4254.55 | 4254.00 | O II | V101 | 0.013 | 0.022 | C |
| 4267.62 | 4267.15 | C II | V6 | 0.325 | 0.522 | A |
| 4276.81 | 4273.10 | O II | V67a | 0.041 | 0.065 | A |
| * | 4275.55 | O II | V67a | * | * | * |
| * | 4275.99 | O II | V67b | * | * | * |
| * | 4276.28 | O II | V67b | * | * | * |
| * | 4276.75 | O II | V67b | * | * | * |
| * | 4277.43 | O II | V67c | * | * | * |
| * | 4277.89 | O II | V67b | * | * | * |
| 4284.98 | 4281.32 | O II | V53b | 0.015 | 0.024 | C |
| * | 4282.96 | O II | V67c | * | * | * |
| * | 4283.73 | O II | V67c | * | * | * |
| * | 4285.69 | O II | V78b | * | * | * |
| 4292.10 | 4291.25 | O II | V55 | 0.021 | 0.032 | B |
| * | 4292.21 | O II | V78c | * | * | * |
| 4295.59 | 4294.78 | O II | V53b | 0.012 | 0.019 | C |
| * | 4294.92 | O II | V53b | * | * | * |
| 4304.41 | 4303.82 | O II | V53a | 0.019 | 0.029 | B |
| * | 4303.61 | O II | V65a | * | * | * |
| 4318.43 | 4317.14 | O II | V2 | 0.042 | 0.065 | A |
| * | 4317.70 | O II | V53a | * | * | * |
| * | 4319.63 | O II | V2 | * | * | * |
| 4326.38 | 4325.76 | O II | V2 | 0.012 | 0.019 | C |
| 4340.94 | 4340.47 | H 5 | H 5 | 30.76 | 46.86 | A |
| 4348.95 | 4349.43 | O II | V2 | 0.094 | 0.141 | A |
| 4363.84 | 4363.21 | [O III] | F2 | 5.975 | 8.873 | A |
| 4379.56 | 4379.11 | N III | V18 | 0.041 | 0.060 | B |
| 4388.41 | 4387.93 | He I | V51 | 0.447 | 0.651 | A |
| 4392.12 | 4391.99 | Ne II | V55e | 0.022 | 0.032 | C |
| * | 4392.00 | Ne II | V55e | * | * | * |
| 4409.76 | 4409.30 | Ne II | V55e | 0.011 | 0.016 | D |
| 4416.01 | 4413.22 | Ne II | V65 | 0.063 | 0.090 | A |
| * | 4413.11 | Ne II | V57c | * | * | * |
| * | 4413.11 | Ne II | V65 | * | * | * |
| * | 4414.90 | O II | V5 | * | * | * |
| * | 4416.97 | O II | V5 | * | * | * |
| 4428.86 | 4428.64 | Ne II | V60c | 0.007 | 0.010 | E |
| * | 4428.52 | Ne II | V61b | * | * | * |
| 4431.06 | 4430.94 | Ne II | V61a | 0.004 | 0.005 | F |
| 4433.07 | 4431.82 | N II | V55a | 0.004 | 0.006 | F |
| * | 4432.74 | N II | V55a | * | * | * |
| * | 4433.48 | N II | V55b | * | * | * |
| 4438.02 | 4437.55 | He I | V50 | 0.048 | 0.067 | B |
| 4453.26 | 4452.36 | O II | V5 | 0.004 | 0.006 | F |
| 4457.66 | 4457.05 | Ne II | V61d | 0.006 | 0.008 | E |
| * | 4457.24 | Ne II | V61d | * | * | * |
| 4471.96 | 4471.49 | He I | V14 | 3.907 | 5.319 | A |
| 4481.47 | 4481.21 | Mg II ? | V4 | 0.019 | 0.026 | C |
| 4488.56 | 4487.72 | O II | V104 | 0.009 | 0.012 | E |
| * | 4488.20 | O II | V104 | * | * | * |
| * | 4489.49 | O II | V86b | * | * | * |
| 4491.76 | 4491.23 | O II | V86a | 0.007 | 0.009 | E |
| * | 4491.07 | C II | | * | * | * |
| 4511.25 | 4510.91 | N III | V3 | 0.032 | 0.042 | C |
| 4515.81 | 4514.86 | N III | V3 | 0.009 | 0.012 | E |
| 4518.77 | 4518.15 | N III | V3 | 0.018 | 0.024 | C |
| 4523.84 | 4523.58 | N III | V3 | 0.014 | 0.018 | D |
| 4530.65 | 4530.41 | N II | V58b | 0.018 | 0.023 | C |
| * | 4530.86 | N III | V3 | * | * | * |
| 4535.32 | 4534.58 | N III | V3 | 0.011 | 0.014 | E |
| 4541.71 | 4541.59 | He II | 4.9 | 0.008 | 0.011 | E |
| 4545.30 | 4544.85 | N III | V12 | 0.015 | 0.020 | D |
| 4553.76 | 4552.53 | N II | V58a | 0.012 | 0.016 | D |
| 4563.07 | 4562.60 | Mg I] | | 0.020 | 0.025 | C |
| 4571.58 | 4571.10 | Mg I] | | 0.329 | 0.414 | A |
| 4591.53 | 4590.97 | O II | V15 | 0.036 | 0.044 | B |
| 4596.64 | 4595.96 | O II | V15 | 0.028 | 0.035 | B |
| * | 4596.18 | O II | V15 | * | * | * |
| 4601.04 | 4601.48 | N II | V5 | 0.083 | 0.103 | A |
| * | 4602.13 | O II | V92b | * | * | * |
| 4614.04 | 4613.14 | O II | V92b | 0.010 | 0.013 | C |
| * | 4613.68 | O II | V92b | * | * | * |
| * | 4613.87 | N II | V5 | * | * | * |
| 4620.90 | 4621.39 | N II | V5 | 0.037 | 0.045 | B |
| 4634.50 | 4634.14 | N III | V2 | 0.220 | 0.264 | A |
| 4641.27 | 4640.64 | N III | V2 | 0.599 | 0.715 | A |
| * | 4641.81 | O II | V1 | * | * | * |
| * | 4641.84 | N III | V2 | * | * | * |
| * | 4643.08 | N II | V5 | * | * | * |
| 4649.78 | 4647.42 | C III | V1 | 0.376 | 0.446 | A |
| * | 4649.13 | O II | V1 | * | * | * |
| * | 4650.25 | C III | V1 | * | * | * |
| * | 4650.84 | O II | V1 | * | * | * |
| 4658.42 | 4658.26 | [Fe III] | F3 | 0.086 | 0.101 | A |
| 4662.04 | 4661.68 | O II | V1 | 0.056 | 0.065 | A |
| 4669.65 | 4669.27 | O II | V89b | 0.005 | 0.005 | E |
| 4676.91 | 4676.24 | O II | V1 | 0.058 | 0.067 | A |
| 4686.16 | 4685.68 | He II | 3.4 | 0.397 | 0.457 | A |
| 4701.73 | 4701.62 | [Fe III] | F3 | 0.016 | 0.018 | C |
| 4706.19 | 4705.35 | O II | V25 | 0.003 | 0.004 | F |
| 4712.08 | 4711.37 | [Ar IV] | F1 | 1.486 | 1.680 | A |
| * | 4713.17 | He I | V12 | * | * | * |
| 4740.62 | 4740.17 | [Ar IV] | F1 | 1.797 | 1.989 | A |
| 4773.28 | 4772.93 | Ne II | | 0.004 | 0.004 | F |
| 4788.95 | 4788.13 | N II | V20 | 0.007 | 0.007 | E |
| 4803.08 | 4802.23 | C II | | 0.017 | 0.017 | C |
| * | 4803.29 | N II | V20 | * | * | * |
| 4815.02 | 4815.55 | S II | V9 | 0.004 | 0.005 | F |
| 4861.98 | 4861.33 | H 4 | H 4 | 100.0 | 100.0 | A |
| 4881.27 | 4881.11 | [Fe III] | F2 | 0.026 | 0.026 | C |
| 4890.74 | 4890.86 | O II | V28 | 0.008 | 0.008 | E |
| 4906.94 | 4906.83 | O II | V28 | 0.019 | 0.019 | E |
| 4922.32 | 4921.93 | He I | V48 | 1.119 | 1.086 | A |
| 4931.54 | 4931.80 | [O III] | F1 | 0.115 | 0.111 | B |
| 4959.54 | 4958.91 | [O III] | F1 | 435.4 | 411.6 | A |

**NGC 6572 S**

| λ Obs. | λ Lab | Ident. | Multiplet | Obs. Flux | Corr. Flux | Error |
|---|---|---|---|---|---|---|
| 4169.84 | 4168.97 | He I | V52 | 0.035 | 0.062 | A |
| * | 4169.22 | O II | V19 | * | * | * |
| 4186.97 | 4185.45 | O II | V36 | 0.036 | 0.063 | A |
| * | 4186.90 | C III | V18 | * | * | * |
| 4200.70 | 4199.83 | He II | 4.11 | 0.025 | 0.044 | B |
| * | 4200.10 | N III | V6 | * | * | * |
| 4220.70 | 4219.37 | Ne II | V52a | 0.012 | 0.021 | C |
| * | 4219.74 | Ne II | V52a | * | * | * |
| 4238.20 | 4236.91 | N II | V48a | 0.008 | 0.013 | C |
| * | 4237.05 | N II | V48b | * | * | * |
| 4242.82 | 4241.24 | N II | V48a | 0.015 | 0.024 | C |
| * | 4241.78 | N II | V48b | * | * | * |
| 4254.68 | 4254.00 | O II | V101 | 0.011 | 0.018 | D |
| 4267.90 | 4267.15 | C II | V6 | 0.302 | 0.491 | A |
| 4277.07 | 4273.10 | O II | V67a | 0.038 | 0.061 | B |
| * | 4275.55 | O II | V67a | * | * | * |
| * | 4275.99 | O II | V67b | * | * | * |
| * | 4276.28 | O II | V67b | * | * | * |
| * | 4276.75 | O II | V67b | * | * | * |
| * | 4277.43 | O II | V67c | * | * | * |
| * | 4277.89 | O II | V67b | * | * | * |
| 4285.06 | 4281.32 | O II | V53b | 0.012 | 0.019 | D |
| * | 4282.96 | O II | V67c | * | * | * |
| * | 4283.73 | O II | V67c | * | * | * |
| * | 4285.69 | O II | V78b | * | * | * |
| 4292.37 | 4291.25 | O II | V55 | 0.014 | 0.022 | C |
| * | 4292.21 | O II | V78c | * | * | * |
| 4295.76 | 4294.78 | O II | V53b | 0.014 | 0.023 | C |
| * | 4294.92 | O II | V53b | * | * | * |
| 4304.46 | 4303.82 | O II | V53a | 0.017 | 0.027 | C |
| * | 4303.61 | O II | V65a | * | * | * |
| 4318.64 | 4317.14 | O II | V2 | 0.042 | 0.065 | B |
| * | 4317.70 | O II | V53a | * | * | * |
| * | 4319.63 | O II | V2 | * | * | * |
| 4326.60 | 4325.76 | O II | V2 | 0.012 | 0.018 | D |
| 4341.18 | 4340.47 | H 5 | H 5 | 30.49 | 46.87 | A |
| 4363.83 | 4363.21 | [O III] | F2 | 5.988 | 8.969 | A |
| 4379.84 | 4379.11 | N III | V18 | 0.037 | 0.055 | B |
| 4388.67 | 4387.93 | He I | V51 | 0.455 | 0.667 | A |
| 4392.28 | 4391.99 | Ne II | V55e | 0.014 | 0.021 | C |
| * | 4392.00 | Ne II | V55e | * | * | * |
| 4410.25 | 4409.30 | Ne II | V55e | 0.013 | 0.019 | C |
| 4416.01 | 4413.22 | Ne II | V65 | 0.061 | 0.087 | A |
| * | 4413.11 | Ne II | V57c | * | * | * |
| * | 4413.11 | Ne II | V65 | * | * | * |
| * | 4414.90 | O II | V5 | * | * | * |
| * | 4416.97 | O II | V5 | * | * | * |
| 4438.30 | 4437.55 | He I | V50 | 0.056 | 0.080 | B |
| 4448.83 | 4448.19 | O II | V35 | 0.006 | 0.009 | F |
| 4453.41 | 4452.36 | O II | V5 | 0.005 | 0.007 | F |
| 4458.35 | 4457.05 | Ne II | V61d | 0.007 | 0.009 | F |
| * | 4457.24 | Ne II | V61d | * | * | * |
| 4472.26 | 4471.49 | He I | V14 | 4.207 | 5.765 | A |
| 4482.14 | 4481.21 | Mg II ? | V4 | 0.016 | 0.022 | D |
| 4489.13 | 4487.72 | O II | V104 | 0.011 | 0.014 | D |
| * | 4488.20 | O II | V104 | * | * | * |
| * | 4489.49 | O II | V86b | * | * | * |
| 4492.02 | 4491.23 | O II | V86a | 0.010 | 0.014 | D |
| * | 4491.07 | C II | | * | * | * |
| 4511.76 | 4510.91 | N III | V3 | 0.032 | 0.043 | B |
| 4515.48 | 4514.86 | N III | V3 | 0.010 | 0.013 | D |
| 4519.31 | 4518.15 | N III | V3 | 0.019 | 0.025 | C |
| 4524.18 | 4523.58 | N III | V3 | 0.016 | 0.021 | C |
| 4530.99 | 4530.41 | N II | V58b | 0.020 | 0.026 | C |
| * | 4530.86 | N III | V3 | * | * | * |
| 4535.75 | 4534.58 | N III | V3 | 0.011 | 0.014 | D |
| 4542.56 | 4541.59 | He II | 4.9 | 0.009 | 0.011 | E |
| 4545.86 | 4544.85 | N III | V12 | 0.013 | 0.016 | D |
| 4554.14 | 4552.53 | N II | V58a | 0.014 | 0.018 | D |
| 4563.21 | 4562.60 | Mg I] | | 0.026 | 0.033 | C |
| 4571.78 | 4571.10 | Mg I] | | 0.430 | 0.545 | A |
| 4591.86 | 4590.97 | O II | V15 | 0.039 | 0.048 | B |
| 4597.00 | 4595.96 | O II | V15 | 0.028 | 0.035 | C |
| * | 4596.18 | O II | V15 | * | * | * |
| 4607.24 | 4607.16 | N II | V5 | 0.091 | 0.112 | A |
| * | 4609.44 | O II | V92a | * | * | * |
| 4614.50 | 4613.14 | O II | V92b | 0.009 | 0.011 | E |
| * | 4613.68 | O II | V92b | * | * | * |
| * | 4613.87 | N II | V5 | * | * | * |
| 4621.41 | 4621.39 | N II | V5 | 0.038 | 0.047 | B |
| 4631.24 | 4630.54 | N II | V5 | 0.028 | 0.034 | C |
| 4634.97 | 4634.14 | N III | V2 | 0.261 | 0.314 | A |
| 4641.69 | 4640.64 | N III | V2 | 0.702 | 0.841 | A |
| * | 4641.81 | O II | V1 | * | * | * |
| * | 4641.84 | N III | V2 | * | * | * |
| * | 4643.08 | N II | V5 | * | * | * |
| 4650.14 | 4647.42 | C III | V1 | 0.423 | 0.503 | A |
| * | 4649.13 | O II | V1 | * | * | * |
| * | 4650.25 | C III | V1 | * | * | * |
| * | 4650.84 | O II | V1 | * | * | * |
| 4658.94 | 4658.26 | [Fe III] | F3 | 0.106 | 0.125 | B |
| 4662.39 | 4661.68 | O II | V1 | 0.066 | 0.078 | B |
| 4670.25 | 4669.27 | O II | V89b | 0.006 | 0.007 | F |
| 4677.01 | 4676.24 | O II | V1 | 0.047 | 0.055 | C |
| 4699.92 | 4699.22 | O II | V25 | 0.004 | 0.005 | F |
| 4706.32 | 4705.35 | O II | V25 | 0.004 | 0.005 | F |
| 4711.87 | 4711.37 | [Ar IV] | F1 | 1.796 | 2.037 | A |
| * | 4713.17 | He I | V12 | * | * | * |
| 4741.05 | 4740.17 | [Ar IV] | F1 | 2.134 | 2.368 | A |
| 4789.24 | 4788.13 | N II | V20 | 0.013 | 0.014 | D |
| 4803.33 | 4802.23 | C II | | 0.021 | 0.023 | C |
| * | 4803.29 | N II | V20 | * | * | * |
| 4861.98 | 4861.33 | H 4 | H 4 | 100.0 | 100.0 | A |
| 4881.83 | 4881.11 | [Fe III] | F2 | 0.044 | 0.044 | C |
| 4891.34 | 4890.86 | O II | V28 | 0.013 | 0.013 | E |
| 4907.35 | 4906.83 | O II | V28 | 0.029 | 0.028 | D |
| 4922.67 | 4921.93 | He I | V48 | 1.524 | 1.478 | A |
| 4931.91 | 4931.80 | [O III] | F1 | 0.161 | 0.155 | B |
| 4959.54 | 4958.91 | [O III] | F1 | 435.4 | 411.1 | A |

**NGC 6790**

| λ Obs. | λ Lab | Ident. | Multiplet | Obs. Flux | Corr. Flux | Error |
|---|---|---|---|---|---|---|
| 4169.96 | 4168.97 | He I | V52 | 0.056 | 0.069 | C |
| * | 4169.22 | O II | V19 | * | * | * |
| 4188.18 | 4185.45 | O II | V36 | 0.141 | 0.172 | B |
| * | 4186.90 | C III | V18 | * | * | * |
| 4201.14 | 4200.10 | N III | V6 | 0.078 | 0.095 | C |
| 4268.52 | 4267.15 | C II | V6 | 0.369 | 0.439 | A |
| 4277.86 | 4273.10 | O II | V67a | 0.043 | 0.051 | B |
| * | 4275.55 | O II | V67a | * | * | * |
| * | 4275.99 | O II | V67b | * | * | * |
| * | 4276.28 | O II | V67b | * | * | * |
| * | 4276.75 | O II | V67b | * | * | * |
| * | 4277.43 | O II | V67c | * | * | * |
| * | 4277.89 | O II | V67b | * | * | * |
| 4283.70 | 4281.32 | O II | V53b | 0.011 | 0.013 | E |
| * | 4282.96 | O II | V67c | * | * | * |
| * | 4283.73 | O II | V67c | * | * | * |
| * | 4285.69 | O II | V78b | * | * | * |
| 4293.68 | 4291.25 | O II | V55 | 0.007 | 0.008 | F |
| * | 4292.21 | O II | V78c | * | * | * |
| 4295.98 | 4294.78 | O II | V53b | 0.008 | 0.010 | E |
| * | 4294.92 | O II | V53b | * | * | * |
| 4304.83 | 4303.82 | O II | V53a | 0.022 | 0.025 | C |
| * | 4303.61 | O II | V65a | * | * | * |
| 4341.79 | 4340.47 | H 5 | H 5 | 40.39 | 47.12 | A |
| 4350.99 | 4349.43 | O II | V2 | 0.024 | 0.028 | C |
| 4364.53 | 4363.21 | [O III] | F2 | 17.69 | 20.45 | A |
| 4389.30 | 4387.93 | He I | V51 | 0.599 | 0.687 | A |
| * | 4391.99 | Ne II | V55e | * | * | * |
| * | 4392.00 | Ne II | V55e | * | * | * |
| 4416.94 | 4413.22 | Ne II | V65 | 0.063 | 0.072 | B |
| * | 4413.11 | Ne II | V57c | * | * | * |
| * | 4413.11 | Ne II | V65 | * | * | * |
| * | 4414.90 | O II | V5 | * | * | * |
| * | 4416.97 | O II | V5 | * | * | * |
| 4438.78 | 4437.55 | He I | V50 | 0.068 | 0.077 | B |
| 4449.56 | 4448.19 | O II | V35 | 0.006 | 0.007 | F |
| 4472.91 | 4471.49 | He I | V14 | 5.295 | 5.926 | A |
| 4516.81 | 4514.86 | N III | V3 | 0.031 | 0.035 | D |
| 4543.03 | 4541.59 | He II | 4.9 | 0.108 | 0.118 | B |
| * | 4544.85 | N III | V12 | * | * | * |
| 4564.39 | 4562.60 | Mg I] | | 0.012 | 0.014 | E |
| 4572.51 | 4571.10 | Mg I] | | 0.291 | 0.317 | A |
| 4592.49 | 4590.97 | O II | V15 | 0.033 | 0.036 | C |
| 4597.28 | 4595.96 | O II | V15 | 0.017 | 0.019 | D |
| * | 4596.18 | O II | V15 | * | * | * |
| 4621.73 | 4621.39 | N II | V5 | 0.035 | 0.038 | D |
| 4635.58 | 4634.14 | N III | V2 | 0.256 | 0.274 | A |
| 4642.24 | 4640.64 | N III | V2 | 0.661 | 0.705 | A |
| * | 4641.81 | O II | V1 | * | * | * |
| * | 4641.84 | N III | V2 | * | * | * |
| * | 4643.08 | N II | V5 | * | * | * |
| 4650.40 | 4647.42 | C III | V1 | 0.612 | 0.652 | A |
| * | 4649.13 | O II | V1 | * | * | * |
| * | 4650.25 | C III | V1 | * | * | * |
| * | 4650.84 | O II | V1 | * | * | * |
| 4659.46 | 4658.26 | [Fe III] | F3 | 0.092 | 0.097 | B |
| 4662.94 | 4661.68 | O II | V1 | 0.049 | 0.052 | C |
| 4677.65 | 4676.24 | O II | V1 | 0.025 | 0.026 | D |
| 4687.15 | 4685.68 | He II | 3.4 | 3.423 | 3.606 | A |
| 4702.94 | 4701.62 | [Fe III] | F3 | 0.024 | 0.025 | D |
| 4706.75 | 4705.35 | O II | V25 | 0.006 | 0.006 | F |
| 4712.03 | 4711.37 | [Ar IV] | F1 | 1.913 | 2.001 | A |
| * | 4713.17 | He I | V12 | * | * | * |
| 4725.65 | 4724.15 | [Ne IV] | F1 | 0.015 | 0.016 | E |
| * | 4725.62 | [Ne IV] | F1 | * | * | * |
| 4741.61 | 4740.17 | [Ar IV] | F1 | 2.679 | 2.781 | A |
| 4862.65 | 4861.33 | H 4 | H 4 | 100.0 | 100.0 | A |
| 4882.39 | 4881.11 | [Fe III] | F2 | 0.037 | 0.037 | D |
| 4891.68 | 4890.86 | O II | V28 | 0.005 | 0.005 | F |
| 4908.28 | 4906.83 | O II | V28 | 0.013 | 0.013 | F |
| 4923.30 | 4921.93 | He I | V48 | 1.551 | 1.534 | A |
| 4932.56 | 4931.80 | [O III] | F1 | 0.191 | 0.188 | B |
| 4960.23 | 4958.91 | [O III] | F1 | 496.8 | 486.6 | A |

**NGC 7027**

| λ Obs. | λ Lab | Ident. | Multiplet | Obs. Flux | Corr. Flux | Error |
|---|---|---|---|---|---|---|
| 4186.71 | 4185.45 | O II | V36 | 0.160 | 0.241 | A |
| * | 4186.90 | C III | V18 | * | * | * |
| 4199.72 | 4199.83 | He II | 4.11 | 0.620 | 0.926 | A |
| * | 4200.10 | N III | V6 | * | * | * |
| 4227.36 | 4227.74 | N II | V33 | 0.147 | 0.216 | B |
| * | 4227.20 | [Fe V] | F2 | * | * | * |
| 4237.72 | 4236.91 | N II | V48a | 0.018 | 0.026 | E |
| * | 4237.05 | N II | V48b | * | * | * |
| 4243.18 | 4241.24 | N II | V48a | 0.036 | 0.052 | C |
| * | 4241.78 | N II | V48b | * | * | * |
| 4250.60 | 4250.65 | Ne II | V52b | 0.010 | 0.014 | F |
| 4253.52 | 4254.00 | O II | V101 | 0.016 | 0.023 | E |
| 4266.97 | 4267.15 | C II | V6 | 0.419 | 0.599 | A |
| 4276.28 | 4273.10 | O II | V67a | 0.046 | 0.065 | C |
| * | 4275.55 | O II | V67a | * | * | * |
| * | 4275.99 | O II | V67b | * | * | * |
| * | 4276.28 | O II | V67b | * | * | * |
| * | 4276.75 | O II | V67b | * | * | * |
| * | 4277.43 | O II | V67c | * | * | * |
| * | 4277.89 | O II | V67b | * | * | * |
| 4286.82 | 4285.69 | O II | V78b | 0.017 | 0.024 | E |
| 4291.27 | 4291.25 | O II | V55 | 0.010 | 0.014 | F |
| * | 4292.21 | O II | V78c | * | * | * |
| 4295.17 | 4294.78 | O II | V53b | 0.012 | 0.017 | F |
| * | 4294.92 | O II | V53b | * | * | * |
| 4303.85 | 4303.82 | O II | V53a | 0.021 | 0.030 | E |
| * | 4303.61 | O II | V65a | * | * | * |
| 4318.09 | 4317.14 | O II | V2 | 0.035 | 0.048 | D |
| * | 4317.70 | O II | V53a | * | * | * |
| * | 4319.63 | O II | V2 | * | * | * |
| 4325.75 | 4325.76 | O II | V2 | 0.016 | 0.021 | F |
| 4340.22 | 4340.47 | H 5 | H 5 | 34.39 | 47.24 | A |
| 4349.61 | 4349.43 | O II | V2 | 0.023 | 0.031 | D |
| 4362.97 | 4363.21 | [O III] | F2 | 19.64 | 26.48 | A |
| 4378.89 | 4379.11 | N III | V18 | 0.098 | 0.131 | B |
| 4387.67 | 4387.93 | He I | V51 | 0.304 | 0.403 | A |
| 4414.93 | 4413.22 | Ne II | V65 | 0.078 | 0.102 | B |
| * | 4413.11 | Ne II | V57c | * | * | * |
| * | 4413.11 | Ne II | V65 | * | * | * |
| * | 4414.90 | O II | V5 | * | * | * |
| * | 4416.97 | O II | V5 | * | * | * |
| 4428.41 | 4428.64 | Ne II | V60c | 0.012 | 0.015 | E |
| * | 4428.52 | Ne II | V61b | * | * | * |
| 4437.23 | 4437.55 | He I | V50 | 0.045 | 0.058 | C |
| 4447.60 | 4448.19 | O II | V35 | 0.024 | 0.030 | D |
| 4452.81 | 4452.36 | O II | V5 | 0.044 | 0.056 | C |
| 4458.28 | 4457.05 | Ne II | V61d | 0.030 | 0.038 | C |
| * | 4457.24 | Ne II | V61d | * | * | * |
| 4471.24 | 4471.49 | He I | V14 | 2.796 | 3.531 | A |
| 4481.26 | 4481.21 | Mg II | V4 | 0.015 | 0.019 | E |
| 4488.14 | 4487.72 | O II | V104 | 0.009 | 0.011 | F |
| * | 4488.20 | O II | V104 | * | * | * |
| * | 4489.49 | O II | V86b | * | * | * |
| 4491.22 | 4491.23 | O II | V86a | 0.009 | 0.011 | F |
| * | 4491.07 | C II | | * | * | * |
| 4510.42 | 4510.91 | N III | V3 | 0.044 | 0.054 | C |
| 4514.51 | 4514.86 | N III | V3 | 0.030 | 0.037 | C |
| 4517.94 | 4518.15 | N III | V3 | 0.038 | 0.047 | C |
| 4523.21 | 4523.58 | N III | V3 | 0.028 | 0.035 | C |
| 4541.34 | 4541.59 | He II | 4.9 | 1.335 | 1.617 | A |
| 4553.24 | 4552.53 | N II | V58a | 0.015 | 0.018 | F |
| 4562.11 | 4562.60 | Mg I] | | 0.024 | 0.029 | E |
| 4570.81 | 4571.10 | Mg I] | | 0.692 | 0.824 | A |
| 4590.87 | 4590.97 | O II | V15 | 0.037 | 0.043 | C |
| 4595.59 | 4595.96 | O II | V15 | 0.025 | 0.029 | D |
| * | 4596.18 | O II | V15 | * | * | * |
| 4606.34 | 4607.16 | N II | V5 | 0.063 | 0.074 | B |
| * | 4609.44 | O II | V92a | * | * | * |
| 4633.81 | 4634.14 | N III | V2 | 1.335 | 1.532 | A |
| 4640.45 | 4640.64 | N III | V2 | 2.851 | 3.259 | A |
| * | 4641.81 | O II | V1 | * | * | * |
| * | 4641.84 | N III | V2 | * | * | * |
| * | 4643.08 | N II | V5 | * | * | * |
| 4647.31 | 4647.42 | C III | V1 | 0.400 | 0.455 | A |
| 4649.83 | 4649.13 | O II | V1 | 0.282 | 0.321 | A |
| * | 4650.25 | C III | V1 | * | * | * |
| * | 4650.84 | O II | V1 | * | * | * |
| 4657.94 | 4658.26 | [Fe III] | F3 | 0.647 | 0.732 | A |
| 4668.76 | 4669.27 | O II | V89b | 0.029 | 0.032 | D |
| 4676.00 | 4676.24 | O II | V1 | 0.103 | 0.116 | C |
| 4685.43 | 4685.68 | He II | 3.4 | 41.90 | 46.68 | A |
| 4701.25 | 4701.62 | [Fe III] | F3 | 0.062 | 0.068 | D |
| 4712.00 | 4711.37 | [Ar IV] | F1 | 3.788 | 4.157 | B |
| * | 4713.17 | He I | V12 | * | * | * |
| * | 4714.30 | [Ne IV] | F1 | * | * | * |
| * | 4717.70 | [Ne IV] | F1 | * | * | * |
| 4724.50 | 4724.15 | [Ne IV] | F1 | 1.538 | 1.676 | A |
| * | 4725.62 | [Ne IV] | F1 | * | * | * |
| 4739.93 | 4740.17 | [Ar IV] | F1 | 7.421 | 8.017 | A |
| 4802.20 | 4802.23 | C II | | 0.020 | 0.021 | E |
| * | 4803.29 | N II | V20 | * | * | * |
| 4814.33 | 4815.55 | S II | V9 | 0.013 | 0.014 | F |
| 4861.01 | 4861.33 | H 4 | H 4 | 100.0 | 100.0 | A |
| 4880.58 | 4881.11 | [Fe III] | F2 | 0.079 | 0.079 | C |
| 4921.63 | 4921.93 | He I | V48 | 0.878 | 0.858 | A |
| 4930.78 | 4931.80 | [O III] | F1 | 0.196 | 0.191 | B |
| 4958.59 | 4958.91 | [O III] | F1 | 500.1 | 479.6 | A |

**TABLE 3**

**Nebular Diagnostics**

| Nebula | [Ar IV] Electron Density ($cm^{-3}$)[a] | Electron Density ($cm^{-3}$) | [O III] Electron Temperature (K)[a] | Electron Temperature (K) |
|---|---|---|---|---|
| IC 4593 | … | 2800[b] | 8380±100 | 8400[b] |
| NGC 6210 | 6520±870 | 5130[c] | 9610±140 | 9690[h] |
| NGC 6543 | 5020±800 | 5000[d] | 8030±100 | 7800[d] |
| NGC 6543S | 4540±740 | … | 7800±100 | … |
| NGC 6572 | 31200±2900 | 38000[e] | 10200±140 | 10600[e] |
| NGC 6572S | 25000±2400 | … | 10280±150 | … |
| NGC 6790 | 74000±9000 | 53000[f] | 12660±240 | 10800[f] |
| NGC 7027 | 84100±12100 | 94000[g] | 14130±290 | 13600[g] |

[a] This Paper

[b] Bohigas & Olguin 1996; $n_e$ from [S II] 6717/6731; temperature from [O III] 4363/5007

[c] Stanghellini & Kaler 1989; $n_e$ from [Ar IV] 4711/4740

[d] Bernard-Salas et al. 2003; $n_e$ is an average of [S III] 18.7μ/33.5μ and [Cl III] 5538/5518; $T_e$ from [O III] 4363/5007

[e] Hyung, Aller & Feibelman 1994; $n_e$ from [Ar IV] 4711/4740; $T_e$ from [O III] 4363/5007

[f] Aller, Hyung & Feibelman 1996; $n_e$ from [Ar IV] 4711/4740; $T_e$ from [O III] 4363/5007

[g] Bernard-Salas et al. 2001; $n_e$ from [Ar IV] 4711/4740; $T_e$ from [O III] 4363/5007

[h] Kaler 1986; $T_e$ from [O III] 4363/5007

**TABLE 4**

**Abundances from Optical Recombination Lines (RL's)**

| Nebula | He I (λ4471) | He II (λ4686) | C II (λ4267) |
|---|---|---|---|
| IC 4593 | 8.40E-02±0.30 | 1.12E-03±0.05 | 1.86E-04±0.49 |
| NGC 6210 | 9.09E-02±0.30 | 1.24E-03±0.04 | 3.38E-04±0.11 |
| NGC 6543 | 10.7E-02±0.40 | Stellar Emission | 6.31E-04±0.25 |
| NGC 6543S | 10.6E-02±0.40 | Stellar Emission | 5.93E-04±0.25 |
| NGC 6572 | 9.30E-02±0.26 | 3.62E-04±0.12 | 5.03E-04±0.12 |
| NGC 6572S | 10.1E-02±0.27 | Stellar Emission | 4.74E-04±0.12 |
| NGC 6790 | 10.4E-02±0.35 | 2.85E-03±0.10 | 4.24E-04±0.14 |
| NGC 7027 | 6.17E-02±0.18 | 3.69E-02±0.12 | 5.78E-04±0.20 |

**TABLE 5**

**O II abundances by wavelength**

| Wavelength | Multiplet(s) | IC4593 | N6210 | N6543 | N6543S | N6572 | N6572S | N6790 | N7027 |
|---|---|---|---|---|---|---|---|---|---|
| 4273.10-4277.89 | 67a, 67b, 67c | 9.33E-04 | 9.25E-04 | 8.27E-04 | 7.11E-04 | 4.20E-04 | 3.99E-04 | 3.34E-04 | 4.23E-04 |
| 4281.32-4285.69 | 53b, 67c, 78b | 1.01E-03 | 7.42E-04 | 1.22E-03 | 1.21E-03 | 3.70E-04 | 2.90E-04 | 1.97E-04 | 3.79E-04 |
| 4291.25-4292.21 | 55, 78c | 1.68E-03 | 1.12E-03 | 1.12E-03 | 7.04E-04 | 1.05E-03 | 7.23E-04 | 2.64E-04 | 4.46E-04 |
| 4294.78-4294.92 | 53b | 1.04E-03 | 1.18E-03 | 1.54E-03 | 8.13E-04 | 5.61E-04 | 6.54E-04 | 2.77E-04 | 4.94E-04 |
| 4303.61-4303.82 | 53a,65a | 1.60E-03 | 1.39E-03 | 1.55E-03 | 1.06E-03 | 4.98E-04 | 4.58E-04 | 4.32E-04 | 5.09E-04 |
| 4315.39-4319.63 | 63c, 78b, 2, 53a | 8.62E-04 | 7.48E-04 | 9.14E-04 | 8.68E-04 | 3.91E-04 | 3.94E-04 | … | 2.92E-04 |
| 4325.76 | 2 | … | 1.21E-03 | 1.41E-03 | 1.56E-03 | 1.26E-03 | 1.20E-03 | … | 1.43E-03 |
| 4487.72-4489.49 | 104, 86b | … | 1.22E-03 | 1.81E-03 | 1.53E-03 | 7.76E-04 | 9.09E-04 | … | 7.17E-04 |
| 4491.23 | 86a | 3.26E-03 | 1.59E-03 | 2.26E-03 | 1.71E-03 | 5.57E-04 | 8.46E-04 | … | 6.53E-04 |
| 4590.97-4596.18 | 15 | 2.68E-03 | 3.23E-03 | 7.35E-03 | 3.72E-03 | 2.40E-03 | 2.52E-03 | 1.64E-03 | 2.15E-03 |
| 4661.63 | 1 | 8.31E-04 | 1.10E-03 | 1.82E-03 | 1.38E-03 | 4.95E-04 | 5.93E-04 | 3.91E-04 | … |
| 4676.24 | 1 | 9.62E-04 | 9.91E-04 | 9.02E-04 | 1.21E-03 | 5.10E-04 | 4.15E-04 | 1.98E-04 | 8.79E-04 |
| 4696.35 | 1 | … | 1.12E-03 | … | … | … | … | … | … |
| 4699.22 | 25 | … | 2.40E-03 | … | … | … | … | … | … |
| 4705.35 | 25 | … | 1.28E-03 | 1.71E-03 | … | 3.31E-04 | 4.28E-04 | … | … |
| 4890.86 | 28 | 3.22E-03 | 1.05E-03 | 2.56E-03 | 1.89E-03 | 6.34E-04 | 1.09E-03 | … | … |
| 4906.33 | 28 | 6.17E-03 | 1.72E-03 | 2.89E-03 | 2.25E-03 | 4.34E-04 | 1.09E-03 | 5.11E-04 | … |
| AVERAGE | | **1.29E-03** | **1.10E-03** | **1.48E-03** | **1.21E-03** | **5.49E-04** | **5.91E-04** | **3.50E-04** | **6.38E-04** |
| Associated Error | | 1.18E-04 | 2.32E-05 | 6.14E-05 | 4.48E-05 | 1.15E-05 | 1.90E-05 | 2.27E-05 | 5.47E-05 |

**Table 6**

**Abundances from Collisionally Excited Lines (CEL's)**

| Nebula | [O III] | C III] |
|---|---|---|
| IC 4593 | 3.59E-04±0.14 | 5.97E-05[a] |
| NGC 6210 | 4.34E-04±0.17 | 1.27E-04[a] |
| NGC 6543 | 5.40E-04±0.20 | 1.89E-04[b] |
| NGC 6543S | 5.84E-04±0.22 | … |
| NGC 6572 | 3.98E-04±0.16 | 5.50E-04[c] |
| NGC 6572S | 3.85E-04±0.15 | … |
| NGC 6790 | 2.43E-04±0.10 | 3.60E-04[d] |
| NGC 7027 | 1.78E-04±0.07 | 2.70E-04[e] |

[a] Kwitter & Henry 1998

[b] Bernard-Salas, Pottasch, Wesselius & Feibelman 2003

[c] Flower & Penn 1981

[d] Aller, Hyung & Feibelman 1996

[e] Kwitter & Henry 1996